\documentclass[%
 amsmath, amssymb, aps,
 reprint,
 superscriptaddress,
 preprintnumbers,
 prd,
nofootinbib,
]{revtex4-2}

\usepackage{orcidlink}
\usepackage[nowatermark]{fixmetodonotes}
\usepackage{placeins} 
\hypersetup{colorlinks = true, allcolors = ., urlcolor = blue}

\usepackage{graphicx}
\usepackage{float}
\usepackage{dcolumn}
\usepackage{bm, amsmath}
\begin{document}

\title{Benchmarking neutrino-nucleus quasielastic scattering model predictions against a missing energy profile obtained using a monoenergetic neutrino beam}

\author{J.~McKean\,\orcidlink{0009-0005-6100-6195}\,}
 \email{mckean.jake.42u@st.kyoto-u.ac.jp}
 \affiliation{Imperial College London, Department of Physics, London SW7 2BZ, United Kingdom}
 \affiliation{Kyoto University, Department of Physics, Kyoto, Japan}
\author{L.~Munteanu\,\orcidlink{0000-0002-2074-8898}\,}
 \email{laura.munteanu@cern.ch}
 \affiliation{European Organization for Nuclear Research (CERN)}
\author{S.~Abe\,\orcidlink{0000-0002-2110-5130}\,}
 \email{seisho@hep.phys.s.u-tokyo.ac.jp}
 \affiliation{Kamioka Observatory, Institute for Cosmic Ray Research, University of Tokyo, Kamioka, Gifu 506-1205, Japan}
 \thanks{Present address: Department of Physics, Graduate School of Science, The University of Tokyo, 7-3-1 Hongo, Bunkyo-ku, Tokyo, 113-0033, Japan}

\date{\today}

\begin{abstract}

We examine three exclusive nuclear ground state shell models implemented in the NEUT neutrino event generator and benchmark them against the recent JSNS$^2$ measurement of missing energy using a monoenergetic neutrino source. The nature of the measurement allows a detailed investigation of nuclear ground-state modeling using a neutrino source, and gives access to a direct measurement of the neutron spectral function in a $^{12}$C nucleus. The NEUT intranuclear cascade and nuclear deexcitation \textsc{NucDeEx} are used to simulate inelastic final-state interactions and nuclear deexcitations respectively. We find that the spectral function (SF) models perform better than relativistic mean field models in modeling both the ground state and the tail of the missing energy distribution when the NEUT cascade and nuclear excitation channels are turned on. We also find that taking into account the missing energy threshold for single nucleon knockout interactions results in all nuclear models being accepted based on the obtained $p$-values. 
\end{abstract}
\maketitle

\section{Introduction}
Precise modeling of neutrino–nucleus interactions is essential for current and future neutrino oscillation experiments, where systematic uncertainties associated with nuclear effects directly impact the interpretation of measured signals~\cite{NuSTEC:2017hzk}. Current and future generation neutrino oscillation experiments heavily rely on Monte Carlo event generators to predict neutrino interactions in the GeV regime. Until recently, neutrino generators mostly relied on inclusive theoretical model implementations that predict differential cross section predictions only as a function of final-state lepton kinematics, even though the generators produce fully exclusive predictions for all particles in the interaction. Over the last few years, generators have undergone an overhaul that accommodates the implementation of more exclusive models by relying on the factorization approach~\cite{Hayato2021,GENIE:2021npt}. This has notably allowed for the implementation of shell-based models such as the Benhar Spectral Function (SF)~\cite{SF:BENHAR1994493,SF:PhysRevD.72.053005,SF:PhysRevC.110.054612}. Recently, the unfactorized energy-dependent relativistic mean field (ED-RMF) model was implemented into the NEUT generator using a tabulated hadron tensor approach~\cite{McKean:2025}. These models have been benchmarked against exclusive electron scattering measurements, and their validation using exclusive neutrino scattering measurements has recently begun~\cite{Gonzalez-Jimenez19, Gonzalez-Jimenez-22,Franco-Patino22, Nikolakopoulos:2022}.

A major challenge in measuring neutrino scattering cross sections arises from the broadband nature of accelerator neutrino beams. The wide energy spectrum complicates the reconstruction of the incoming neutrino energy and smears out final-state observables, making it difficult to disentangle the modeling of the nuclear ground state from nuclear effects. An exception to this comes from neutrinos produced by kaon decay at rest (KDAR). In this process, a kaon undergoes a two-body decay producing a neutrino ($K^+\rightarrow\mu^+\nu_\mu$) with a well defined energy of 235.5\,MeV, with a branching ratio of 63.6\%~\cite{PDG:10.1093/ptep/ptaa104}.  \par

Recently, the JSNS$^2$ collaboration has released its first measurement of the missing energy spectrum using neutrinos from a KDAR source on a carbon target~\cite{KDAR:Marzec:2025}. This measurement can be used to study the neutron spectral function of carbon, which is a widely used nuclear target in experiments (for example, by experiments such as T2K~\cite{Abe:2011:T2K}, MINERvA~\cite{MINERvA:2013zvz}, and NOvA~\cite{NOvA:2007rmc}). These experiments operate at energies of $\approx1$\,GeV, so KDAR neutrinos are comparable to the regions of interest for accelerator-based experiments (such as T2K and NOvA, and in the future, DUNE~\cite{DUNE:PhysRevD.105.072006} and Hyper-K~\cite{Hyper-Kamiokande:2025fci}). Moreover, these energies are particularly relevant for testing the approximations on which neutrino interaction generators are built. For example, many generators rely on the plane wave impulse approximation (PWIA), which assumes that the neutrino scatters off a single isolated nucleon and that the outgoing particles propagate as undistorted plane waves, neglecting correlations and interactions with the residual nucleus. The energy region of the KDAR neutrinos is low enough that the energy transfers involved in charged-current (CC) interactions become sensitive to nuclear medium effects (for example, related to the nuclear ground state, Pauli blocking, or nucleon-nucleon correlations), which provides a unique opportunity to probe them directly. 

In this work, we focus on benchmarking three exclusive shell-based models implemented in the NEUT event generator~\cite{Hayato2021} (version 6.0.3) against the JSNS$^2$ measurement of KDAR neutrinos. We choose to focus on the NEUT event generator because, at the time of writing, it offers the widest choice of exclusive shell-based models within a single generator, allowing us to test the models in a consistent framework.

\section{Interaction Models} For neutrinos with an energy of 235.5\,MeV, the vast majority of scatters will take place via charged-current quasielastic (CCQE) processes ($\nu_\mu n \rightarrow \mu^-p$). Multi-nucleon knock-out processes (so-called 2p2h, or npnh, interactions) can happen in this energy regime, but represent a less than 10\% contribution to the total cross section. Higher-order processes such as resonant pion production and deep inelastic scatters are negligible at these energies. The following paragraphs describe in detail the CCQE models that we test in this work. For 2p2h interactions, we use the model by Nieves {\it et al.}~\cite{Nieves2p2h:PhysRevC.83.045501, NIEVES201272,Gran:PhysRevD.88.113007}, which is the default model implemented in NEUT. Resonant pion production is modeled using the Rein-Sehgal model with added lepton mass corrections~\cite{REIN198179, BergerSehgal:PhysRevD.76.113004, GraczykSobczyk:PhysRevD.77.053003}. For a full description of the available models in the NEUT event generator, see~\cite{Hayato2021}.

NEUT offers several CCQE models. Among them, we chose to focus on two versions of the Benhar SF model, and the ED-RMF model. \par
\begin{figure}[htbp]
    \centering
    \includegraphics[width=1.0\columnwidth]{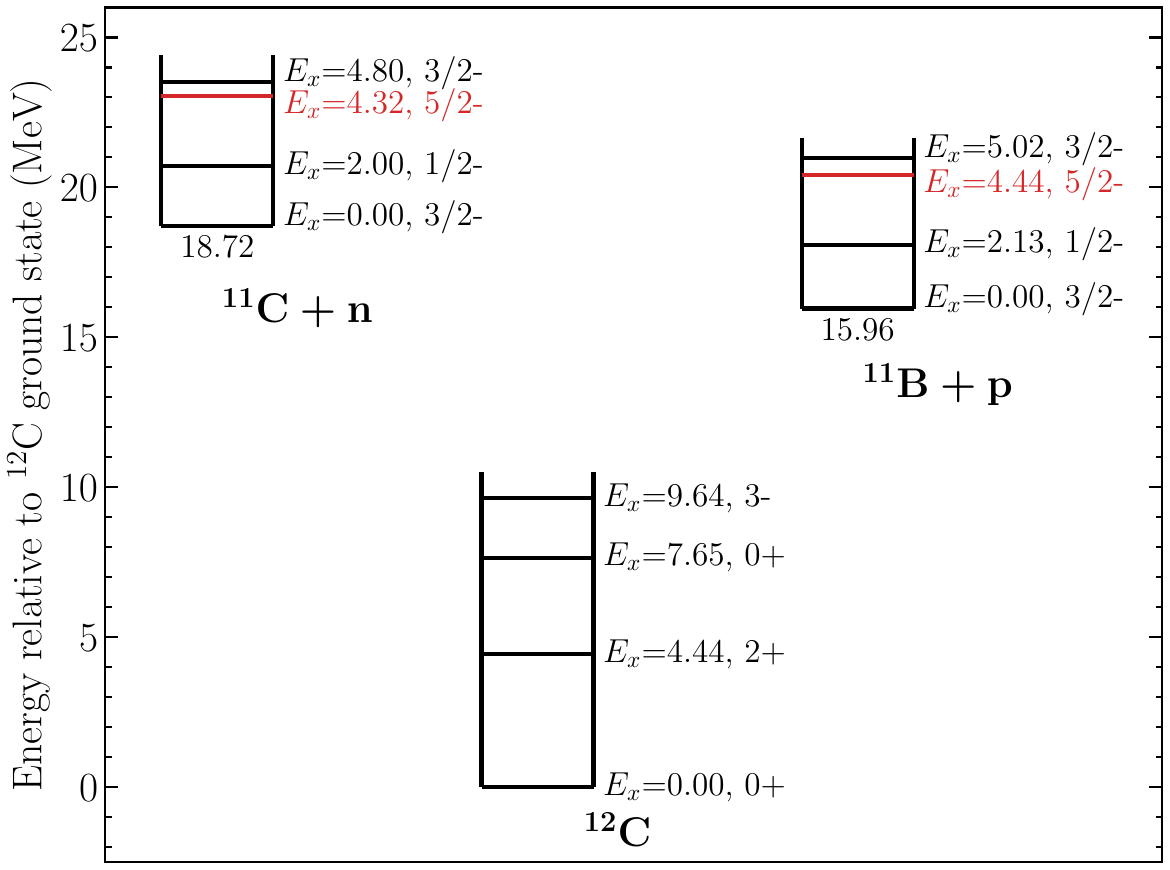}
    \caption{Energy levels of $^{12}$C, $^{11}$B, and $^{11}$C relative to the $^{12}$C ground state based on Refs.~\cite{levelA11:KELLEY201288,levelA12:KELLEY201771}.
             The excited states with $J^\pi=5/2^-$ of $^{11}\mathrm{C}$ and $^{11}\mathrm{B}$, shown in red, are not included in the SF$^*$~\cite{SF:PhysRevC.110.054612} while the other three levels are included.}
    \label{fig:level_diagram}
\end{figure}

The SF model provides the two-dimensional probability $P(\bold{p},\tilde{E})$ of removing a proton with momentum $\bold{p}$ and removal energy $\tilde{E}$~\cite{SF:BENHAR1994493,SF:PhysRevD.72.053005,Abe:2025:PhysRevD.111.033006},
\begin{equation}
    \tilde{E}=E_k-E_{k'}-E_{p'}+M
\end{equation}
where $E_k$, $E_{k'}$, and $E_{p'}$ represent the energy of the incoming lepton, outgoing lepton, and outgoing nucleon, respectively, while $M$ represents the target nucleon mass.
Note that the removal energy is identical to the missing energy discussed here in the absence of final-state interactions (FSI). Recently, a new carbon SF with high removal energy resolution~\cite{SF:PhysRevC.110.054612} has also been implemented in NEUT. This new SF (denoted as SF$^{*}$ in this work) distinguishes between the ground state and the first and second excited states of $^{11}\text{B}^*$ (with excitation energies at 2.13\,MeV and 5.02\,MeV, as shown in Fig.~\ref{fig:level_diagram}) that were smeared out by coarse binning in the previous SF. Thus, it is expected to provide a more precise description of gamma-ray emissions by the nuclear deexcitation. Since the SF tables are obtained from exclusive electron scattering measurements $(e,e'p)$, they provide only the spectral function for protons. To obtain the neutron spectral function, an additional correction is required. While assuming that the momentum distributions of protons and neutrons in symmetric nuclei are the same, a shift to the removal energy distribution is applied to neutrons, following the discussion in Refs.~\cite{Artur:PhysRevC.86.024616,Artur:PhysRevLett.108.052505}.
\begin{equation}
    P_{\text{neutron}}(\bold{p},\tilde{E}) = P_{\text{proton}}(\bold{p},\tilde{E}-\Delta\tilde{E}),
\end{equation}
where $\Delta\tilde{E}=2.76$\,MeV for the carbon nucleus~\cite{isotope:firestone_shirley_baglin_chu_zipkin_1997}. This correction takes into account the fact that neutrons are more deeply bound in the nucleus than protons due to the absence of the Coulomb potential. In the case of SF$^*$, this correction approximately reproduces the removal energy corresponding to the first and second discrete excited states of $^{11}\text{C}^*$ (with excitation energies at 2.00\,MeV and 4.80\,MeV). The results of applying this correction for both SF and SF$^{*}$ will be discussed in this paper.
Pauli blocking is implemented using a simple step function, which requires the momentum of the outgoing nucleon to exceed the Fermi surface $p_F=209$\,MeV~\cite{SF:PhysRevD.72.053005}.
In the high-removal energy and high-momentum region, the SF and SF$^{*}$ models include a non-negligible contribution from short-range correlations (SRCs). In NEUT, events with $\tilde{E}>100\,$MeV or $|\mathbf{p}|>300$\,MeV are defined as originating from SRCs. The SRCs, according to the NEUT definition, account for approximately 10\% of the total integrated SF probability. Their contribution to the CCQE cross section is energy dependent: at the KDAR neutrino energy of 235.5\,MeV it is about 1\%, while at higher neutrino energies it can reach approximately 7\%. For these events, the target nucleon is assumed to exist as part of a correlated nucleon pair in a back-to-back configuration, and the correlated partner nucleon has momentum opposite to that of the initial target nucleon. In such events, both nucleons undergo subsequent FSI.

Unlike the SF approach, the ED-RMF model, which was recently implemented in NEUT as discussed in Ref.~\cite{McKean:2025}, uses an unfactorised treatment. The benefit of this approach is that the initial and final states are treated in a consistent theoretical framework. In this description, the hadron tensor $H^{\mu\nu}_{\kappa}$ for a given nuclear shell, $\kappa$, is given as 
\begin{equation}
H^{\mu\nu}_{\kappa} = \frac{1}{2j+1}\sum_{m_{j}, s} [J^{\mu}_{\kappa, m_{j}, s}]^{*} [J^{\nu}_{\kappa, m_{j}, s}],
\end{equation}
\noindent where $m_j$ and $s$ are the third component of the total angular momentum of the hole state and the spin of the scattered nucleon state, respectively. $j$ denotes the total angular momentum, and $2j+1$ is the total occupancy of a given shell. The hadron current $J^{\mu}_{\kappa, m_{j}, s}$ is described by single-particle RMF states~\cite{Gonzalez-Jimenez19} as 
\begin{equation}
J^{\mu}_{\kappa, m_{j}, s} 
= \int \text{d}\mathbf{p} \hspace{2pt} \bar{\psi}_s(\mathbf{p} + \mathbf{q}, \mathbf{p_{N}}) \hspace{2pt}  \Gamma^{\mu} \hspace{2pt}  \Psi_{\kappa}^{m_{j}}(\mathbf{p}).
\end{equation}
\noindent Here, $\Psi_{\kappa}^{m_{j}}(\mathbf{p})$ is the initial-state RMF wavefunction, $\Gamma^{\mu}$ is the transition operator for the CCQE interaction in the CC2 form~\cite{Kelly:PhysRevC.56.2672}, and $\bar{\psi}_s(\mathbf{p} + \mathbf{q}, \mathbf{p_{N}})$ is the final-state wavefunction that is the solution to the Dirac equation in the presence of a nuclear potential. $\mathbf{p}$ is the momentum of the initial nucleon, $\mathbf{p_{N}}$ is the momentum of the final-state nucleon, and $\mathbf{q}$ is the three momentum transfer. The model has multiple nuclear potentials available: ED-RMF~\cite{Gonzalez-Jimenez19}, energy-dependent $A$-independent relativistic optical potentials (EDAI)~\cite{Cooper-potentials}, the real-only component of the EDAI optical potential (rEDAI) and no potential which simplifies to the relativistic plane wave impulse approximation (RPWIA). The inclusion of the nuclear potential in the final state is not present in the SF models and provides additional information about elastic FSI that the scattered nucleon undergoes. In order to model the nucleus in a similar fashion to the SF model, without the need for factorization, the missing energy profile, which is factorized from the precomputed hadron tensor tables, is modeled using Gaussian nuclear shells~\cite{McKean:2025}; the Gaussian shell structure for carbon is given in Table~\ref{tab:C12-RMF-Em-profile} and a ``background'' shell contribution is included that is modeled in three distinct regions in $E_{m}$. The shell positions are chosen to be solutions to RMF equations or are taken from recent measurements where they exist. The shell occupancies and widths are chosen to mimic the SF~\cite{SF:BENHAR1994493, Benhar08} following the procedure described in Ref.~\cite{Franco-Patino22}. The ED-RMF potential has a natural implementation of Pauli blocking due to the orthogonality of the wavefunctions for the initial and final states~\cite{Gonzalez-Jimenez19}.

\begin{table} [htbp]
    \caption{Occupancies, central values and widths of the Gaussian functions used to model the neutron nuclear shells for the ED-RMF model on $^{12}$C. The values used are taken from Refs.~\cite{Franco-Patino22, McKean:2025}.}
    \label{tab:C12-RMF-Em-profile}
    \centering
    \begin{tabular*}{0.8\columnwidth}{@{\extracolsep{\fill}}c c c c }
    \hline \hline
    Shell & Occupancy & $E_{m}^{\kappa}$ [MeV] & $\sigma^{\kappa}$ [MeV] \\ [0.5 ex]
    \hline
    $1s_{1/2}$ & 1.90/2.00 & 37.00 & 10.00 \\
    $1p_{3/2}$ & 3.30/4.00 & 18.72 & 2.00   \\ [1ex]
    \hline \hline
    \end{tabular*}
\end{table}

\begin{figure}[htbp]
    \centering
    \includegraphics[width=1.0\columnwidth]{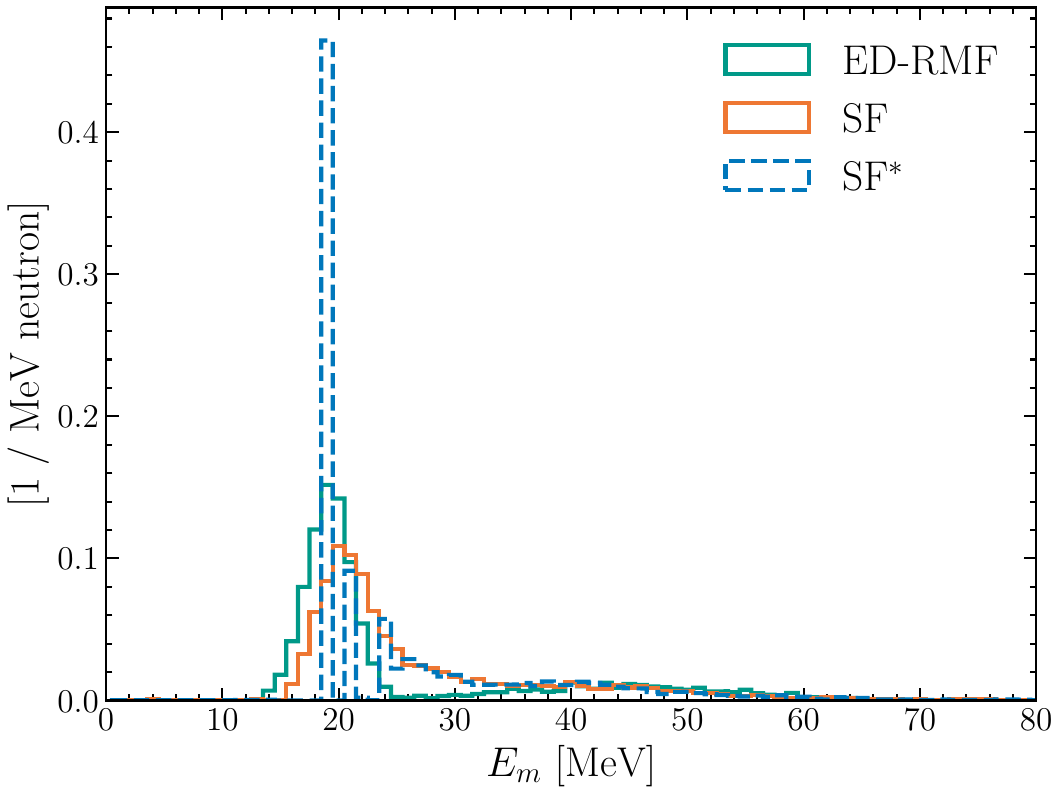}
    \caption{Area-normalised missing energy distributions for the ED-RMF, SF, and SF$^{*}$ $\nu_\mu$ CCQE models in NEUT version 6.0.3. The SF$^{*}$ models the ground state of $^{11}$C with a Dirac delta function at 18.72\,MeV.}
    \label{fig:norm_Em_profiles}
\end{figure}

The area-normalised missing energy distributions for these three models are shown in Figure~\ref{fig:norm_Em_profiles}. The difference between SF and SF$^*$ appears only in the low-missing energy region, while they have identical distribution in $E_m\gtrsim25$\,MeV, as expected. The ED-RMF model shows a dip around 25~MeV which is a result of the Gaussian modeling of the nuclear shells. There is a shift of about 1\,MeV in peak positions between SF and ED-RMF. This is caused by the fact that the SF smears $p$-shells rather than modeling $1p_{3/2}$ by Gaussian like the ED-RMF model. In addition, the leading edge of the ED-RMF missing energy distribution begins at lower $E_{m}$ values than the SF model due to different smearing widths. It is important to note that the missing energy profile is modeled identically for the ED-RMF, EDAI, rEDAI and RPWIA final-state potentials.

The NEUT FSI cascade is based on the mean free path of a particle travelling through the nuclear medium~\cite{Hayato2021}. Particles are tracked and stepped by a fixed size of 0.2~fm through the nuclear medium until it leaves the nucleus. During this process, nucleon rescattering based on the Bertini model~\cite{Bertini:1972vz} and pion absorption based on Ref.~\cite{SALCEDO:1988:557} can occur, leading creation and ejection of other particles from the nucleus. The NEUT FSI cascade is inherently an inelastic process; during nucleon rescatters, both nucleons involved in the interaction are tracked through further reinteractions in the cascade. In contrast, the ED-RMF model incorporates the final-state potential into calculating the kinematics of the final-state nucleon. As a result, elastic FSI is included. The FSI from the NEUT cascade does not pose a double-counting problem~\cite{Nikolakopoulos:2022}. \par

Recently, a dedicated nuclear deexcitation event generator \textsc{NucDeEx}~\cite{Abe:2024:PhysRevD.109.036009,NucDeEx:2v9j-ncf4,Abe:NucDeEx} was implemented in NEUT. The branching ratios of high-energy continuum excited states above the separation energy are calculated with the Hauser-Feshbach model~\cite{HF:PhysRev.87.366}, as implemented in the nuclear reaction code \textsc{TALYS}~\cite{Koning:2023}. NucDeEx simulates emission of various light particles, such as nucleons, deuterons, tritons, $^{3}$He, and alpha particles. On the other hand, the gamma-ray branching ratios from low-lying discrete excited states of $^{11}\text{B}^*$ and $^{11}\text{C}^*$ are taken from experimental measurements~\cite{PANIN2016204}. It should be noted that \textsc{NucDeEx} enables to simulate deexcitation from various nuclei other than single nucleon knockout, $^{11}\text{C}^*$ in this analysis. A detailed description of the implementation in NEUT is given in Ref.~\cite{NucDeEx:2v9j-ncf4}. \par

A single-nucleon knockout interaction can only occur once the missing energy is above the single nucleon knockout (1NKO) threshold. The 1NKO threshold corresponds to the neutron separation energy. The neutron separation energy corresponds to
\begin{equation}
    S_{n} = m(^{11}\text{C}) + m_{n} - m(^{12}\text{C}),
\end{equation}
and has a value of 18.72\,MeV for $^{12}\text{C}$~\cite{KELLEY201771}. The SF model is extracted from coarsely binned electron scattering data and the ED-RMF missing energy modeling parameters are chosen to replicate the SF for carbon. Both the SF and ED-RMF models use a double-sided Gaussian description. As a result of this choice, both of their missing energy distributions extend below the 1NKO, which is unphysical. Due to the broadband nature of neutrino fluxes in most experiments, this issue is effectively unnoticeable once the neutrino flux is integrated over. However, the JSNS$^{2}$ measurement is sensitive enough to highlight the potential inaccuracies in nuclear models such as SF, namely single-nucleon knockout interactions occurring below the 1NKO threshold. To highlight this issue, in the following section the comparisons between the models and the measurements are performed twice -- with and without applying a cutoff in the missing energy distributions at the 1NKO.

\section{Results and discussion} 
\label{sec:results_discussion}
The JSNS$^{2}$ measurement provides a shape-only differential cross-section measurement as a function of missing energy~\cite{KDAR:Marzec:2025}. To compare to this measurement, $\nu_\mu$-$^{12}$C interactions were generated by NEUT version 6.0.3 using a monochromatic neutrino energy of $235.5$\,MeV. The ED-RMF model uses an axial mass of $M_{A}^{\text{QE}} = 1.05\,$GeV/$c$ with a dipole parameterization for the axial-vector form factor and the parameterisation of Ref.~\cite{Kelly:PhysRevC.70.068202} for the vector form factor. The SF and SF$^{*}$ models use an axial mass of $M_{A}^{\text{QE}} = 1.03\,$GeV/$c$ with dipole parameterization and BBBA05~\cite{BBBA05:BRADFORD2006127} for vector form factors. The choice of the axial mass value gives a negligible impact on the results, since this is a shape-only measurement and the dipole parametrization used has mainly a normalization effect. Samples were produced with and without the NEUT intranuclear cascade. Additional effects from nuclear deexcitation were also investigated using \textsc{NucDeEx}. The samples were then processed into a simplified event format using the NUISANCE framework~\cite{Stowell:2017}. \par
 JSNS$^2$~\cite{KDAR:Marzec:2025} measures the missing energy, $E_{m}$, defined as
\begin{equation}
\begin{split}
    E_{m} =~ & E_{\nu_\mu} (235.5\text{ MeV})  - 
m_{\mu} (105.7 \text{ MeV})\\ & +  [m_{n} - m_{p}](1.3 \text{ MeV})- E_{\text{vis}} \\ & = 131.2 \text{~MeV}-E_{\text{vis}}.
\end{split}
\end{equation} 
\noindent where $E_{\nu}(235.5\,\text{MeV})$ is the KDAR neutrino energy, $m_{\mu} (105.7 \text{ MeV})$ is the muon mass and $E_\text{vis}$ is the visible energy inside the detector. This latter quantity has contributions from the muon energy, the kinetic energy of protons, as well as from de-excitation photons. The event selection requires a muon in the final state and that the total visible energy satisfies $20\,\text{MeV} <E_\text{vis}< 150\,\text{MeV}$. \par

In order to compare the generated samples to the measurement, the following proxy for $E_m$ is defined using truth-level information:
\begin{equation}
   E_{m} = E_{\nu}(235.5\,\text{MeV}) - E_{\mu} - \sum T_{p} - E_{\gamma},
\end{equation}

\noindent where $\sum T_{p}$ is the sum of proton kinetic energies and $E_{\gamma}$ is the energy of the gamma rays emitted by nuclear deexcitation of the residual nuclei. To match the signal definition as closely as possible, a muon in the final state is required and the total kinetic energy of the muon and proton(s), $T = T_{\mu} + \sum T_{p}$,  must satisfy $20\,\text{MeV} <T < 150\,\text{MeV}$. It should be noted that the kinetic energy of other charged particles, such as deuterons and alpha particles, is not included in this missing energy calculation. Although NucDeEx can simulate the emission of these particles, their visible energy is expected to be much lower than their kinetic energy due to stronger quenching effect than that of protons. Since the quenching effect for these heavy charged particles is not discussed in Ref.~\cite{KDAR:Marzec:2025}, we did not include their contribution in the missing energy calculation. 

To quantify the agreement between the generator predictions and the measurement, a $\chi^{2}$ analysis is performed using the available cross-section values and covariance matrices that are released with the measurement. The $\chi^{2}$ statistic used is defined as 
\begin{equation*}
    \chi^{2} = \sum_{i,\,j} \big( D - M \big)_{i} \big(\text{Cov}^{-1} \big)_{ij} \big( D - M\big)_{j},
\end{equation*}
\noindent where $D_{i/j}$ and $M_{i/j}$ are the respective measurement and generator values for the bins $i,j$ that are used in the calculation and $\text{Cov}$ is the covariance matrix. Assuming a Gaussian approximation, an associated $p$-value can be obtained for each $\chi^{2}$ and its associated degrees of freedom. We define the case where a model is excluded by the measurement as that in which the corresponding $p$-value given the number of degrees of freedom is below 0.05. \par

Separate covariance matrices are available for the statistical uncertainty, proton Birks' constant uncertainty, and generator systematic uncertainties. During the calculation of the $\chi^{2}$, a selection is applied to the $E_{\text{vis}}$ values such that $45\,\text{MeV} \leq E_{\text{vis}} < 120\,\text{MeV}$, in order to exclude bins with null cross section. The selection is applied to the events and covariance matrices before the $\chi^{2}$ calculation. The $E_{\text{vis}}$ selection corresponds to $4.8\,\text{MeV} \leq E_{m} < 84.8\,\text{MeV}$. The lower bound of this $E_{m}$ selection is below the 1NKO. \\

Computing the $\chi^2$ and corresponding $p$-values in the case where the 1NKO is applied requires the introduction of two modifications to the measurement and associated covariance matrix. Taking this threshold into account, the $E_{\text{vis}}$ selection becomes $18.7\,\text{MeV} \leq E_{\text{vis}} < 84.8\,\text{MeV}$, which removes three degrees of freedom from the $\chi^{2}$ calculation. The corresponding rows and columns from the covariance matrices are also removed. This operation is justified under the assumption of a gaussian distribution of errors, which is inherent to the treatment via a covariance matrix. Since this measurement is presented in the unfolded truth space, it is assumed that possible spill-over artifacts from adjacent bins were corrected for during the unfolding procedure and the binning size was chosen accordingly. Importantly, one of the bins which is removed (likely corresponding to the $1p_{3/2}$ peak) contains the 1NKO threshold itself, so the removal of this bin also removes the $1p_{3/2}$ peak from the $\chi^{2}$ calculation. The bins around the $1p_{3/2}$ peak present large correlations, which can significantly alter the value of the $\chi^{2}$.\par

\begin{figure*}[htbp]
    \centering
    \begin{minipage}{0.32\textwidth}
        \centering
        \includegraphics[width=\linewidth]{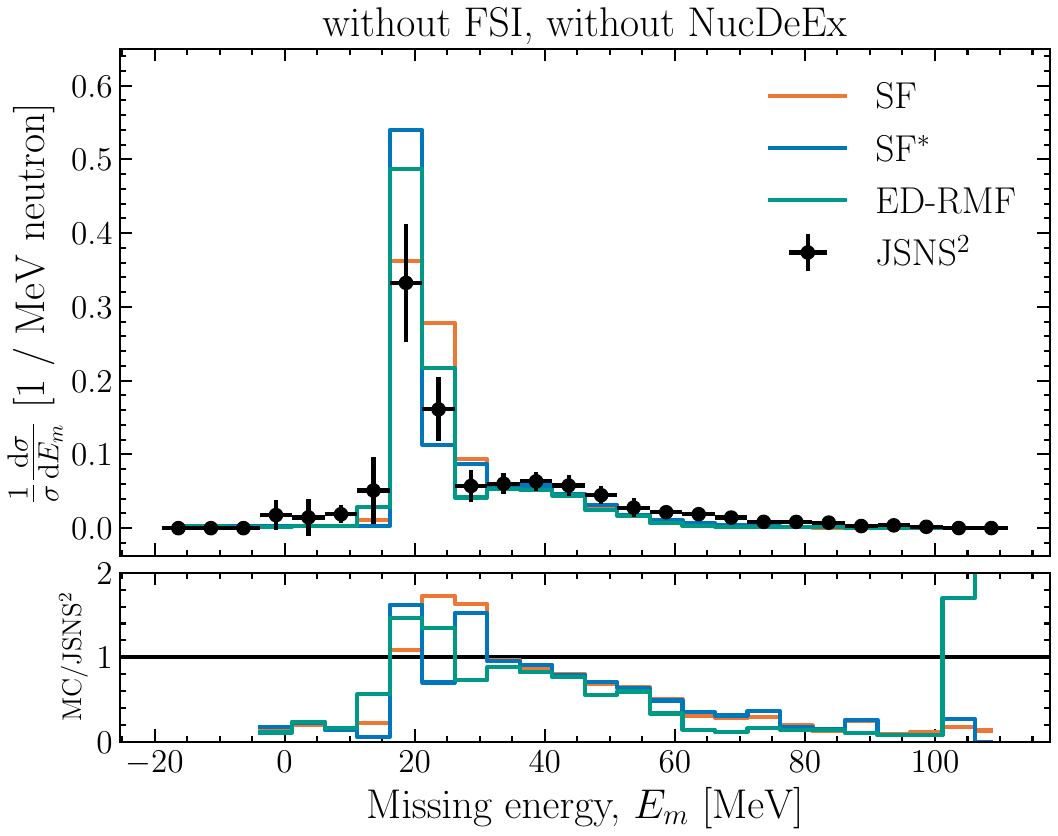}
    \end{minipage}\hfill
    \begin{minipage}{0.32\textwidth}
        \centering
        \includegraphics[width=\linewidth]{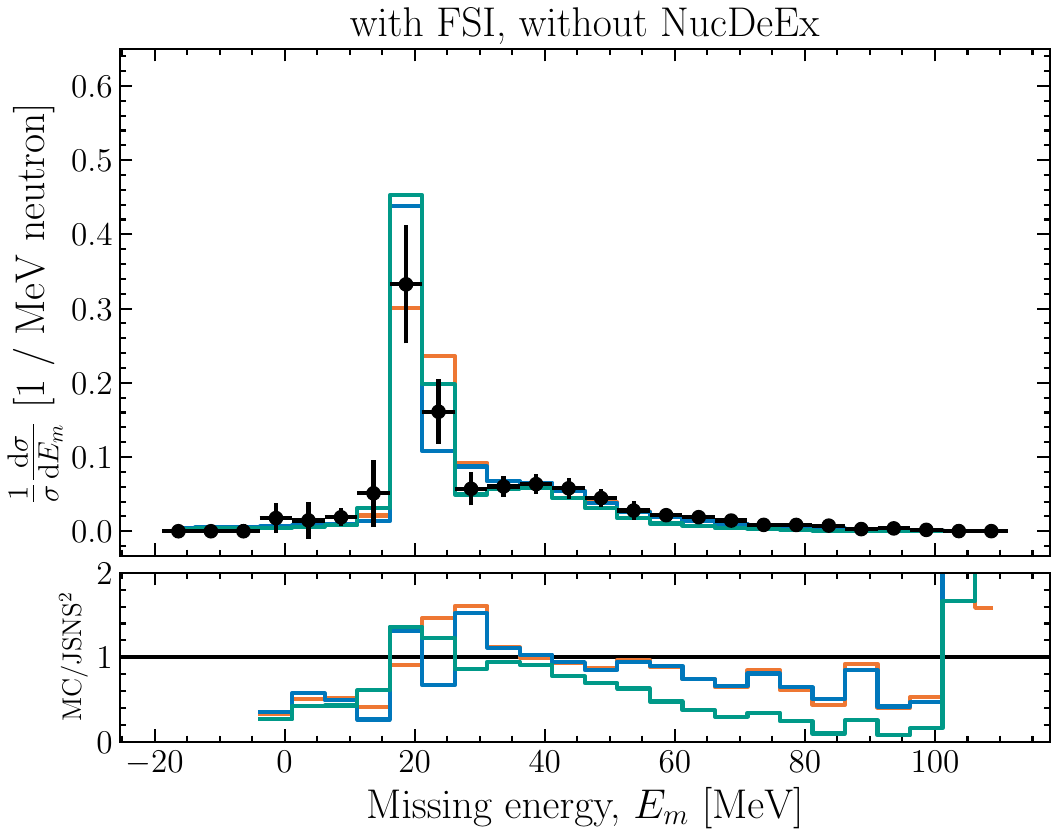}
    \end{minipage}\hfill
    \begin{minipage}{0.32\textwidth}
        \centering
        \includegraphics[width=\linewidth]{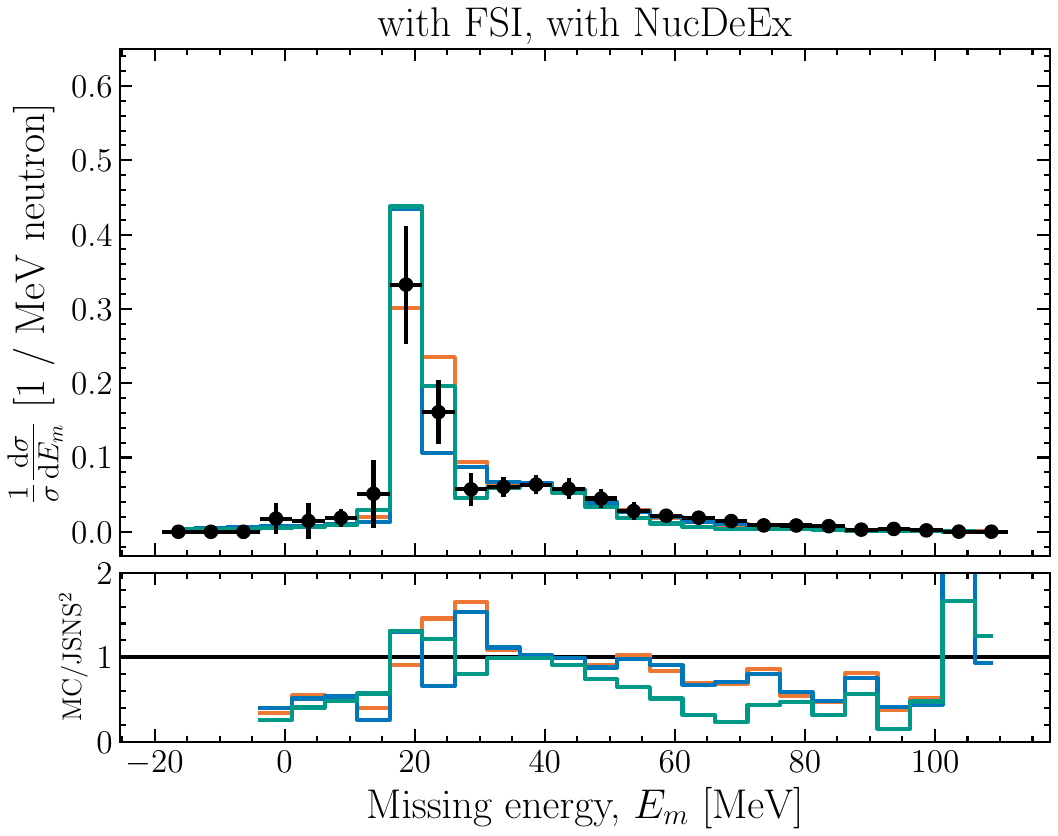}
    \end{minipage}

    \caption{Shape-only missing energy differential cross section for the SF, SF$^{*}$, and ED-RMF models. The figures show cases where the NEUT FSI cascade and \textsc{NucDeEx} deexcitation are switched off (left), the NEUT FSI cascade is on and \textsc{NucDeEx} deexcitation is switched off (center) and both the NEUT FSI cascade and \textsc{NucDeEx} deexcitation are switched on (right). The data points are taken from JSNS$^2$ measurement~\cite{KDAR:Marzec:2025}.}
    \label{fig:MissE}
\end{figure*}

\begin{figure*}[htbp]
    \centering
    \begin{minipage}{0.32\textwidth}
        \centering
        \includegraphics[width=\linewidth]{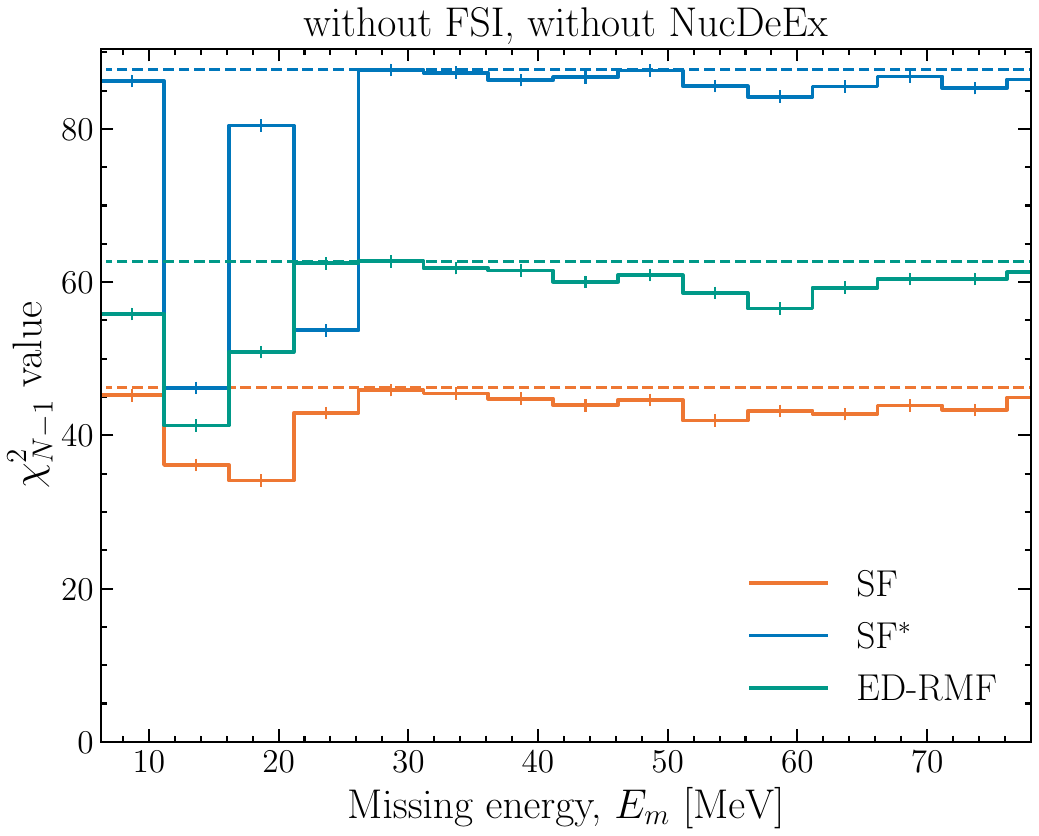}
    \end{minipage}\hfill
    \begin{minipage}{0.32\textwidth}
        \centering
        \includegraphics[width=\linewidth]{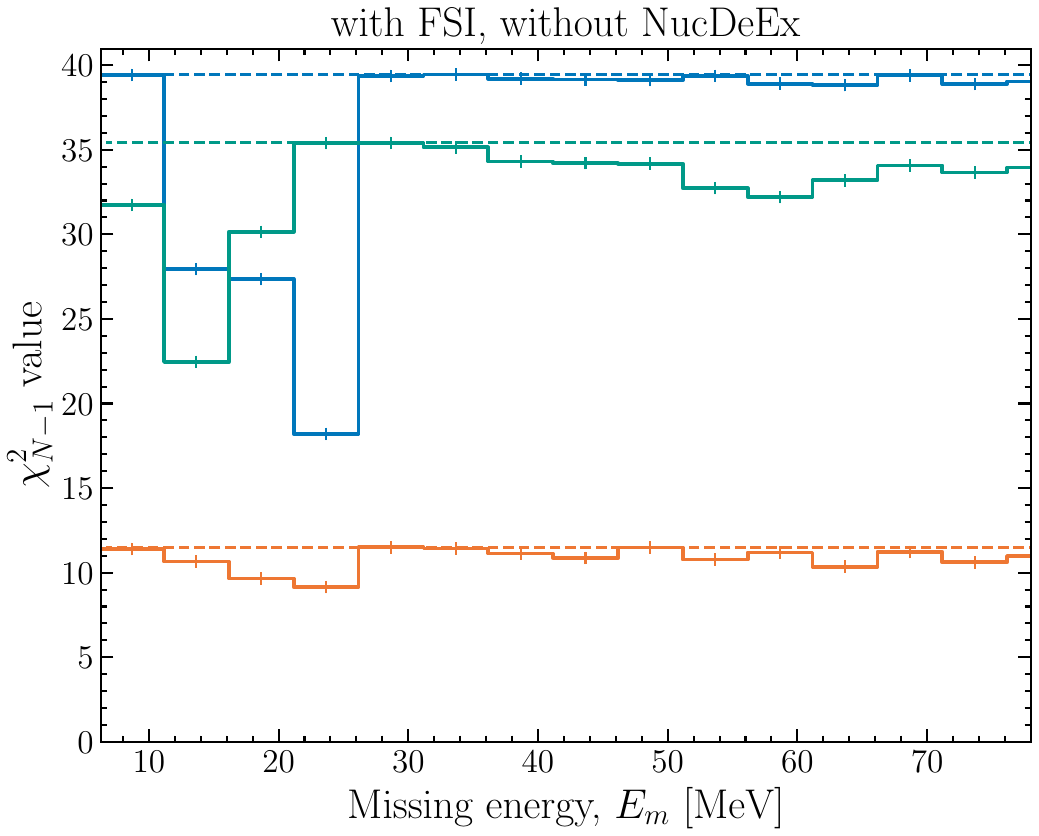}
    \end{minipage}\hfill
    \begin{minipage}{0.32\textwidth}
        \centering
        \includegraphics[width=\linewidth]{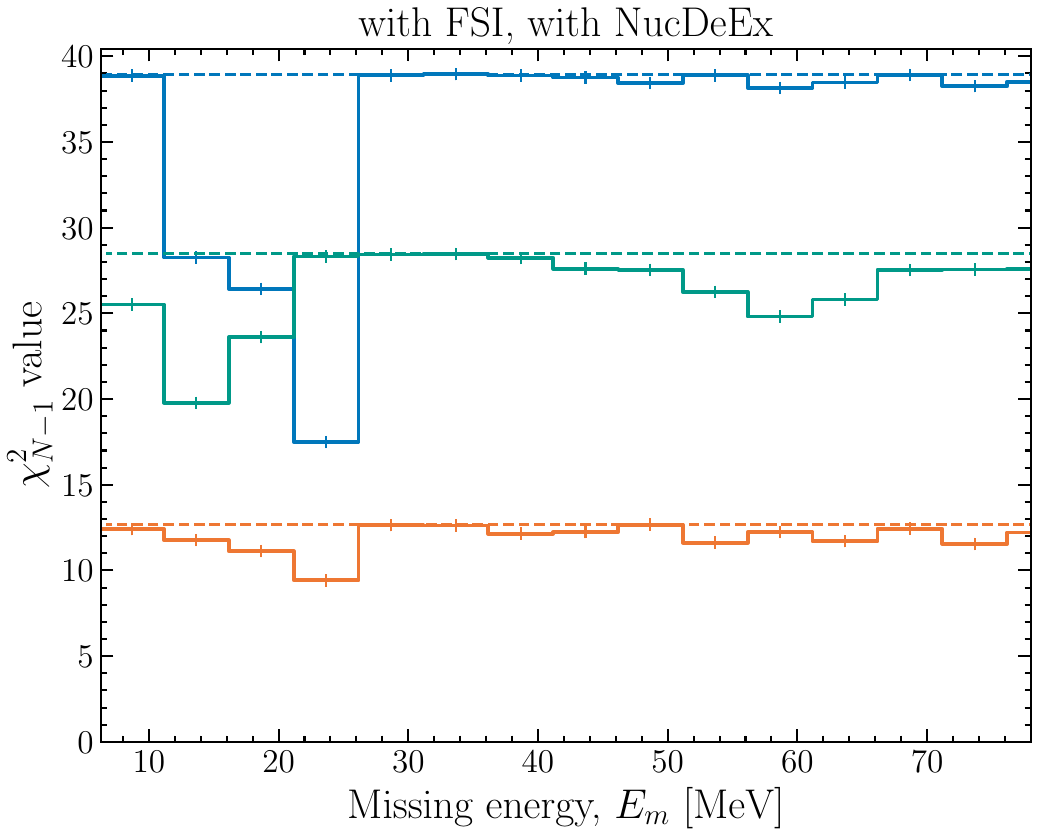}
    \end{minipage}

    \caption{$\chi^{2}_{N-1}$ value without the single nucleon knockout threshold. The figures show cases where the NEUT FSI cascade and \textsc{NucDeEx} deexcitation are switched off (left), the NEUT FSI cascade is on and \textsc{NucDeEx} deexcitation is switched off (center) and both the NEUT FSI cascade and \textsc{NucDeEx} deexcitation are switched on (right). Dashed lines indicate the full $\chi^{2}$ value with all bins.}
    \label{fig:chi2_Nminus1_wo_1NKO}
\end{figure*}

\begin{figure*}[htbp]
        \centering
    \begin{minipage}{0.32\textwidth}
        \centering
        \includegraphics[width=\linewidth]{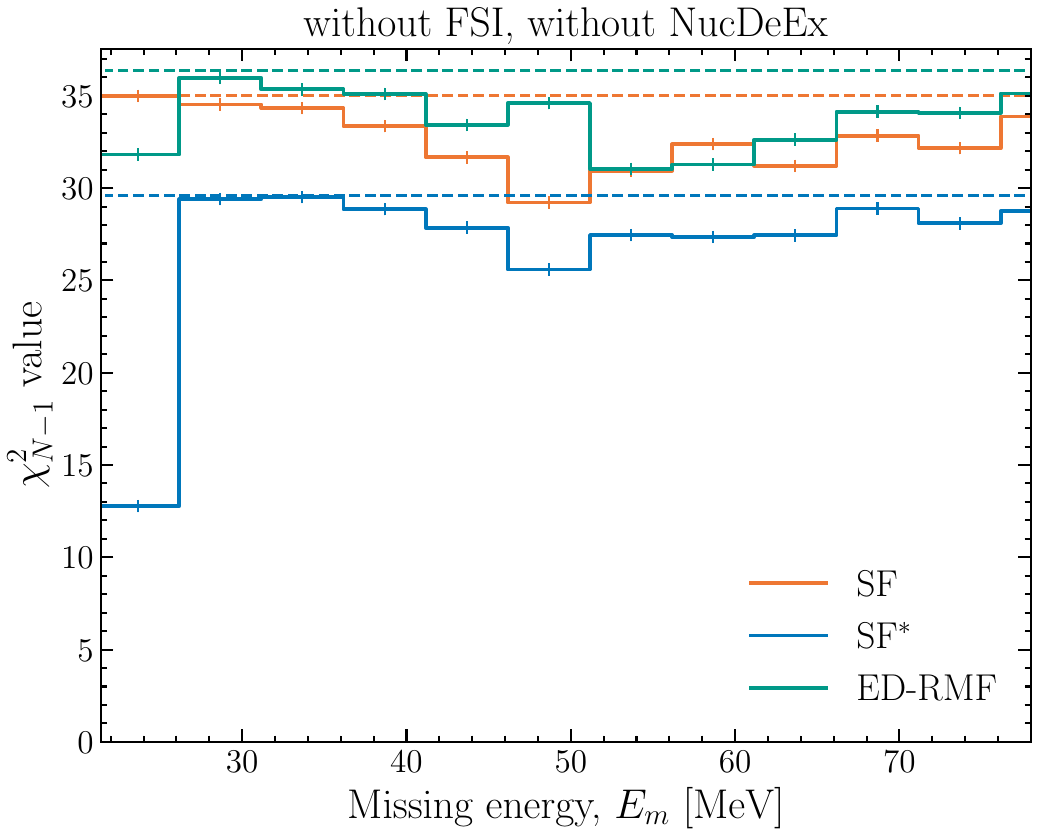}
    \end{minipage}\hfill
    \begin{minipage}{0.32\textwidth}
        \centering
        \includegraphics[width=\linewidth]{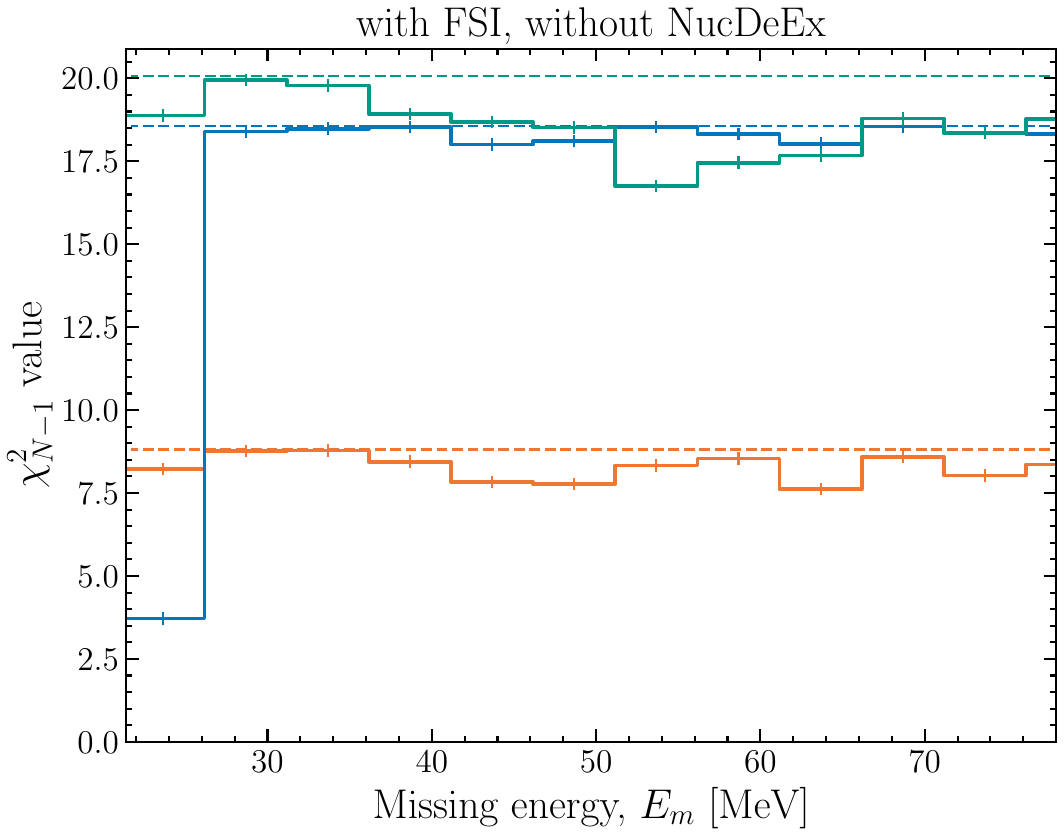}
    \end{minipage}\hfill
    \begin{minipage}{0.32\textwidth}
        \centering
        \includegraphics[width=\linewidth]{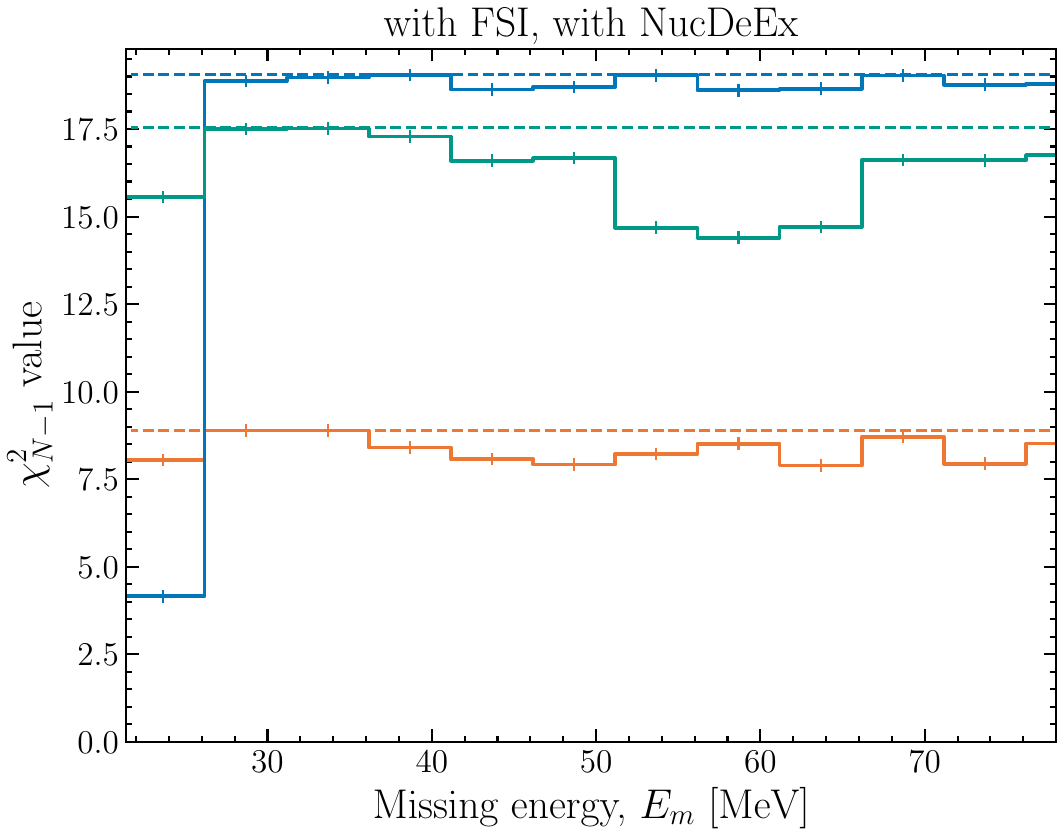}
    \end{minipage}

    \caption{The same as Fig.~\ref{fig:chi2_Nminus1_wo_1NKO} but showing only bins above the single nucleon knockout threshold.}
    \label{fig:chi2_Nminus1_1NKO}
\end{figure*}

In an effort to isolate the effect of the NEUT FSI cascade from the modeling of the ground state and the de-excitation routine, each component is turned on sequentially and compared to the JSNS$^2$ measurement in Figure~\ref{fig:MissE}. The obtained $\chi^{2}$ values, with and without the 1NKO threshold and the corresponding $p$-values, are shown in Table~\ref{tab:model_config_chi2}. To better understand the contribution of each individual bin to the overall $\chi^{2}$ value, a $\chi^{2}_{N-1}$ analysis is shown in Figures~\ref{fig:chi2_Nminus1_wo_1NKO} and \ref{fig:chi2_Nminus1_1NKO} for the cases without and with the 1NKO threshold respectively. In such an analysis, each bin, along with its corresponding covariance matrix row and column, is removed from the total $\chi^{2}$ calculation to isolate its effect (but still accounting for the remaining correlations). 

The models in NEUT, without the NEUT cascade and \textsc{NucDeEx} deexcitation, are compared to the measurement in the left side of Figure~\ref{fig:MissE}.
The SF$^{*}$ and ED-RMF models both overestimate the ground-state $1p_{3/2}$ peak, whereas the SF model reproduces the height of the peak bin adequately. Before the $1p_{3/2}$ peak, the SF$^{*}$ model has no strength, unlike the SF and ED-RMF models, as expected of the Dirac delta description of the ground-state in the SF$^{*}$ model. All models struggle to reproduce the bin after the $1p_{3/2}$ peak, but in different ways -- the SF and ED-RMF models overestimate the contribution, while the SF$^{*}$ model significantly underestimates it. From Figure~\ref{fig:chi2_Nminus1_wo_1NKO}, only the bins adjacent to the ground-state peak contribute most significantly to the overall $\chi^{2}$ for the SF$^{*}$ model. For the ED-RMF and SF models, both the $1p_{3/2}$ peak bin and its adjacent bins contribute most significantly. These observations follow the qualitative trends that can be expected from the removal energy distributions in Figure~\ref{fig:norm_Em_profiles}. When applying the 1NKO threshold, this is no longer the case for the SF and ED-RMF models, where the contribution to the $\chi^{2}$ predominantly comes from the tail of the distribution. For the SF$^{*}$ model the dominant $\chi^{2}$ contribution now comes from the bin just after the $1p_{3/2}$ peak. The second $1s_{1/2}$ peak around 40\,MeV is inadequately described by all models. The lack of inelastic FSI means that the strength of nucleon rescatters that increases the size of the tail of the distribution is not present, as is expected from all three models.  The $\chi^{2}$ values show that all models are rejected in the case where the NEUT cascade and \textsc{NucDeEx} are switched off.\par

Including the effect of the NEUT FSI cascade, without the effect of \textsc{NucDeEx} (shown in the center panel of Figure~\ref{fig:MissE}), shows an improvement of the $\chi^{2}$ values, both before and after applying the 1NKO cutoff, as can be seen in Table~\ref{tab:model_config_chi2}. As a result of the inclusion of the inelastic FSI cascade, all models show a shift of strength from the $1p_{3/2}$ peak towards the tail, which drives the improvement in $\chi^{2}$ values. The $p$-values indicate that the SF$^{*}$ and ED-RMF models are still excluded, but the $p$-value for the SF model now significantly increases to 0.78 without applying the 1NKO threshold. Figure~\ref{fig:chi2_Nminus1_wo_1NKO} (center) shows that with the inclusion of the inelastic NEUT FSI cascade, the dominant contribution from the SF$^{*}$ model comes from the bin after the $1p_{3/2}$ peak, however the qualitative trend stays similar for the SF and ED-RMF models, with an improved overall $\chi^2$. Supported by Figure~\ref{fig:chi2_Nminus1_1NKO}, the tails of the SF$^{*}$ and SF models are affected in a nearly identical way, as expected since they have the same removal energy distribution above about 25\,MeV. However, the dominant contribution to the $\chi^{2}$ for the SF$^{*}$ model is driven by the bin after the $1p_{3/2}$ peak. The ED-RMF model performs worst in the tail of the distribution. No model is quantitatively excluded when applying the 1NKO threshold to this configuration.

The effect of including both the NEUT cascade and \textsc{NucDeEx} deexcitation is shown in the right side of Figure~\ref{fig:MissE}. The ground-state $1p_{3/2}$ peak is well captured by the SF model but the SF$^{*}$ and ED-RMF models still overpredict the measurement. The effect of the deexcitation routine is to slightly shift strength from the tail of the distribution to the $1p_{3/2}$, with a more pronounced effect on the ED-RMF distribution. From Figure~\ref{fig:chi2_Nminus1_1NKO} (left), the effect of the deexcitation routine introduces minor changes to the tails of the SF and SF$^{*}$ models, and an overall reduction in the $\chi^{2}$ for the ED-RMF model. the SF$^{*}$ model with the 1NKO threshold owes more than half of its $\chi^{2}$ value to the first bin in the figure, which corresponds to the bin just above the $1p_{3/2}$ peak. This could indicate a missing source of strength in the transition between the $1p_{3/2}$ and $1s_{1/2}$ peaks. The $\chi^{2}$ values, given in Table~\ref{tab:model_config_chi2}, show that without the 1NKO threshold, only the SF model is accepted and the SF$^{*}$ and ED-RMF models are rejected. Including the 1NKO threshold, all models are accepted based on the $p$-value.

%

\begin{table} [htbp]
    \caption{$\chi^{2}$ and $p$-values for the SF, SF$^{*}$ and RMF nuclear models in NEUT. The NEUT cascade (given by the FSI column), nuclear deexcitation and single nucleon knockout threshold (given by the 1NKO column) is shown for each configuration. Rows highlighted in bold font are not rejected given the $p$-values. The $p$-values are rounded to two significant figures.}
    \label{tab:model_config_chi2}
    \centering
    \begin{tabular*}{1.0\columnwidth}{@{\extracolsep{\fill}}c c c c c c }
    \hline \hline
    CCQE model & FSI & Deexcitation & 1NKO & $\chi^{2}/N_{\text{d.o.f}}$ & $p$-value\\ [0.5 ex]
    \hline
    ED-RMF & No & No & No & 62.72/16 & 0.00 \\
    SF & No & No & No & 46.28/16 &  0.00\\
    SF$^{*}$ & No & No & No & 87.76/16 & 0.00\\
    \hline
    ED-RMF & Yes & No & No &35.41/16 & 0.00\\
    \textbf{SF} & \textbf{Yes} & \textbf{No} & \textbf{No} & \textbf{11.50/16} & \textbf{0.78} \\
    SF$^{*}$ & Yes & No & No &39.45/16 & 0.00\\
    \hline
    ED-RMF & Yes & Yes & No & 28.48/16 & 0.03\\
    \textbf{SF} & \textbf{Yes} & \textbf{Yes} & \textbf{No} &\textbf{12.68/16} & \textbf{0.70}\\
    SF$^{*}$ & Yes & Yes & No & 38.96/16 & 0.00 \\
    \hline
    ED-RMF & No & No & Yes & 36.37/13 & 0.00\\
    SF & No & No & Yes & 35.02/13 & 0.00\\
    SF$^{*}$ & No & No & Yes& 29.60/13 & 0.01\\
    \hline
    \textbf{ED-RMF }& \textbf{Yes} & \textbf{No} & \textbf{Yes} & \textbf{20.07/13} & \textbf{0.09}\\
    \textbf{SF} & \textbf{Yes} & \textbf{No} & \textbf{Yes} & \textbf{8.80/13} & \textbf{0.79}\\
    \textbf{SF}$\mathbf{^{*}}$ & \textbf{Yes} & \textbf{No} & \textbf{Yes} & \textbf{18.56/13} & \textbf{0.14} \\
    \hline
    \textbf{ED-RMF} & \textbf{Yes} & \textbf{Yes} & \textbf{Yes} & \textbf{17.54/13} & \textbf{0.18}\\
    \textbf{SF} & \textbf{Yes} & \textbf{Yes} & \textbf{Yes} & \textbf{8.89/13} & \textbf{0.78}\\
    \textbf{SF}$\mathbf{^{*}}$ & \textbf{Yes} & \textbf{Yes} & \textbf{Yes} & \textbf{19.05/13} & \textbf{0.12} \\
    \hline \hline
    \end{tabular*}
\end{table}

In the discussion so far, ED-RMF is used as the default RMF-based model. Other nucleon potential models, such as RPWIA, EDAI, and rEDAI, lead to only slight changes in the results (see Appendix~\ref{sec:ap:RMF-nuc-pots}). Furthermore, in the ED-RMF model, it is possible to vary the shell occupancies, as well as central values and widths of the Gaussian functions, as shown in Table.~\ref{tab:C12-RMF-Em-profile}. The impact of varying this parameterization is discussed in Appendix~\ref{sec:ap:RMF-Em-profile}.\par

Additionally, the impact of varying the FSI strength in the NEUT cascade is shown in Appendix~\ref{sec:ap:FSI_strength}. While this work uses the NEUT cascade, investigating other generator cascades, such as NuWro~\cite{NuWro} as shown in Ref.~\cite{Franco-Patino26}, is necessary to draw more general conclusions about how FSI affects agreement with the measurement. \par

Finally, there exist contributions to the cross section on $^{12}$C that are not modeled by NEUT or other standard neutrino event generators. One such contribution arises from transitions to bound states of $^{12}$N without direct nucleon knockout, i.e. $^{12}$C$(\nu_\mu, \mu^-)^{12}$N$^*$. This process has been measured at LAMPF~\cite{Koetke:1992yk} at a mean neutrino energy of $202\,\text{MeV}$ and calculated theoretically~\cite{Lee:2024pbl}. Such events could in principle be reconstructed at a fixed $E_m \approx 17$~MeV, and would manifest as an enhancement of the cross section directly at the $1p_{3/2}$ peak. To assess the impact of this unmodeled contribution on the conclusions presented here, the cross section in the $1p_{3/2}$ peak region was artificially increased by 11\% (before normalization), consistent with the LAMPF measurement in Ref.~\cite{Koetke:1992yk}. For the ED-RMF and SF$^{*}$ models, which already overestimate the $1p_{3/2}$ peak, the $\chi^2$ values deteriorate further, reinforcing their rejection by the measurement. For the SF model, the $\chi^2$ increases slightly but the $p$-value remains acceptable and the model is not rejected. The overall conclusions of this study are therefore robust with respect to this effect. Nevertheless, accurate modeling of these bound-state transition contributions in neutrino event generators would be valuable for future precision measurements.

\section{Conclusion}
The recent measurement of the missing energy distribution of $^{12}$C using a monoenergetic neutrino beam by JSNS$^{2}$ was used to benchmark three nuclear models within the NEUT generator. The SF, SF$^{*}$, and ED-RMF nuclear models were tested alongside the impact of the NEUT FSI cascade and \textsc{NucDeEx} deexcitation routines. \par

When including the effects of the NEUT FSI cascade and deexcitation and comparing to the full measurement, the SF model yields the best agreement with the measurement. Despite being more theoretically grounded and robust, the SF$^{*}$ and ED-RMF models are excluded. Upon closer investigation, there are several reasons for this discrepancy. The ED-RMF model uses a double sided Gaussian to parametrize the $1p_{3/2}$ ground-state peak, introducing a non-physical contribution below the 1NKO threshold, which drives the increase in the overall $\chi^{2}$. The ED-RMF model also underpredicts measurement in the bin after the  $1p_{3/2}$ peak. A single-sided Gaussian would be a more natural method to model the ground state. The SF$^{*}$ model does not have a contribution below the 1NKO threshold, but appears to overestimate the overall strength of the $1p_{3/2}$ peak and underestimate the contribution in the transition between the $1p_{3/2}$ and $1s_{1/2}$ peaks.\par

The inclusion of the FSI cascade has the most noticeable impact on the agreement between the measurement and the simulations, both with and without the inclusion of the 1NKO threshold. A natural extension to this work would be an in-depth study of different FSI models from different generators. \par

In the analysis where the 1NKO threshold is applied, the overall sensitivity of the measurement is degraded due to the exclusion of the $1p_{3/2}$ peak, which drives the discriminating power, and its strongly correlated neighbors. This leads to all models being accepted given the measurement once the FSI cascade and the \textsc{NucDeEx} are included. Despite this, a more detailed $\chi^{2}_{N-1}$ analysis allows to identify that $\chi^{2}$ contributions stem from different sources for different models -- the SF$^{*}$ model agreement is still driven by the transition between the $1p_{3/2}$ and $1s_{1/2}$ peaks, whereas the SF and ED-RMF models are now driven by the contribution in the tail. For ED-RMF, in particular, the larger $\chi^{2}$ contribution from the tail compared to SF could indicate a potential mis-modeling of inelastic FSI in NEUT, or indicate an incorrect treatment of SRCs.\par

Although most neutrino scattering measurements are not sensitive to such detailed aspects of the modeling of nuclear effects due to the broadband nature of artificial neutrino fluxes, these comparisons highlight an issue in the modeling of the neutron spectral functions employed in the NEUT event generator. The detailed statistical analysis points to several different sources of discrepancy between different NEUT models and the JSNS$^2$ neutron spectral function measurement, and offers insight into specific ways in which each model can be modified.

\section{Acknowledgments}
J.~McKean was supported by the STFC and UKRI, and Grant-in-Aid for JSPS Research Fellows, JSPS KAKENHI Grant No. 25KF0223. S.~Abe was supported by JSPS KAKENHI Grant No. 23KJ0319 and 25H00631.
The authors thank Stephen Dolan, Raúl González-Jiménez, Kevin McFarland and the T2K Neutrino Interactions Working Group for fruitful discussions and suggestions during the preparation of this manuscript. Additionally, the authors would like to thank Eric Marzec and Joshua Spitz for their support in the exploitation of the measurement data release, as well as Noah Steinberg for his help in verifying the obtained results.
Also, the authors thank Artur Ankowski and Omar Benhar for their helpful discussions on the validation of the implementation of the new spectral function.

\appendix
\section{Comparing final-state nuclear potentials}\label{sec:ap:RMF-nuc-pots}
\begin{figure*}[htbp]
    \centering
        \begin{minipage}{0.32\textwidth}
        \centering
        \includegraphics[width=\linewidth]{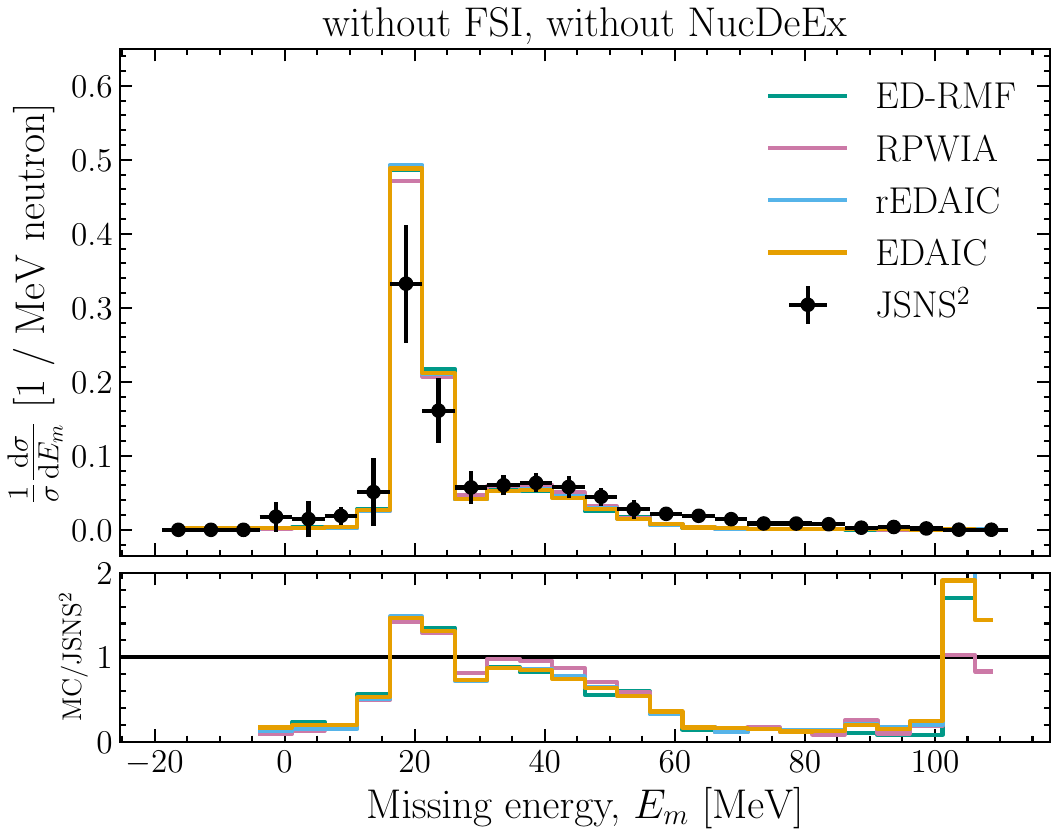}
    \end{minipage}\hfill
    \begin{minipage}{0.32\textwidth}
        \centering
        \includegraphics[width=\linewidth]{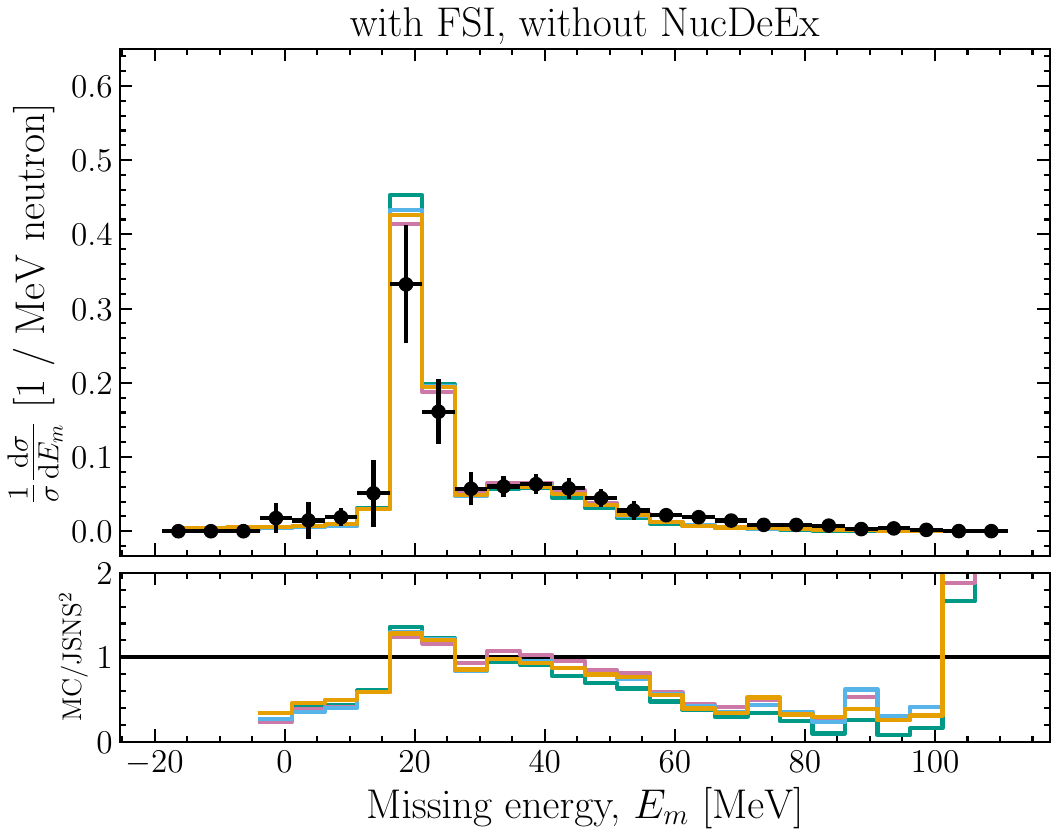}
    \end{minipage}\hfill
    \begin{minipage}{0.32\textwidth}
        \centering
        \includegraphics[width=\linewidth]{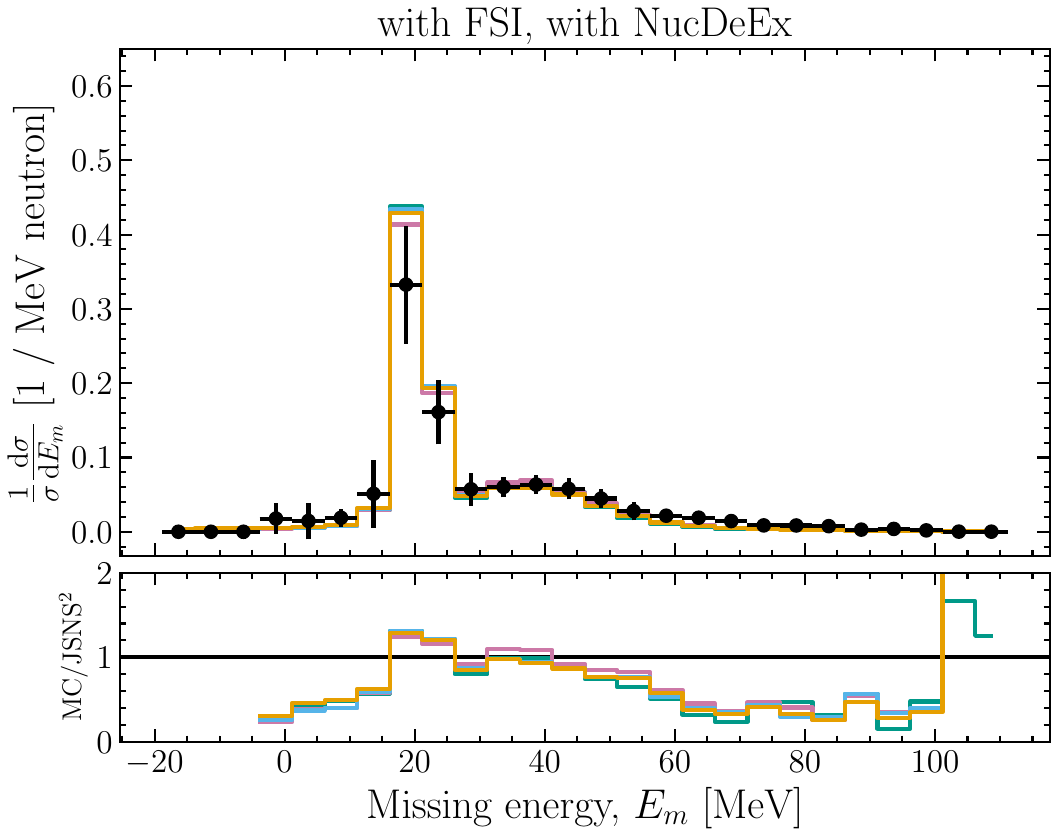}
    \end{minipage}
    \caption{The same as Figure~\ref{fig:MissE} but showing the ED-RMF, RPWIA, EDAI and rEDAI nuclear potentials. The data points are taken from the JSNS$^{2}$ measurement~\cite{KDAR:Marzec:2025}.}
    \label{fig:missE_NucPots}
\end{figure*}

The RMF-based models in NEUT allows the user to change the final-state nuclear potential between the ED-RMF, RPWIA, EDAI and rEDAI potentials. All nuclear potentials available were compared against the same JSNS$^{2}$ measurement and a similar $\chi^{2}_{N-1}$ analysis was performed. The associated $\chi^{2}$ and $p$-values are given in Table~\ref{tab:DiffPots_Em_profile_config_chi2}. From Figure~\ref{fig:missE_NucPots}, we see that all nuclear potentials show a similar distribution. Because the RMF-based models use the same missing energy parameterisation for all nuclear potentials, and because the measurement is of a shape-only cross-section, the distributions are expected to be almost identical. However, differences between the models are present and highlighted below.\par
\begin{figure*}[htbp]
    \centering
    \begin{minipage}{0.32\textwidth}
        \centering
        \includegraphics[width=\linewidth]{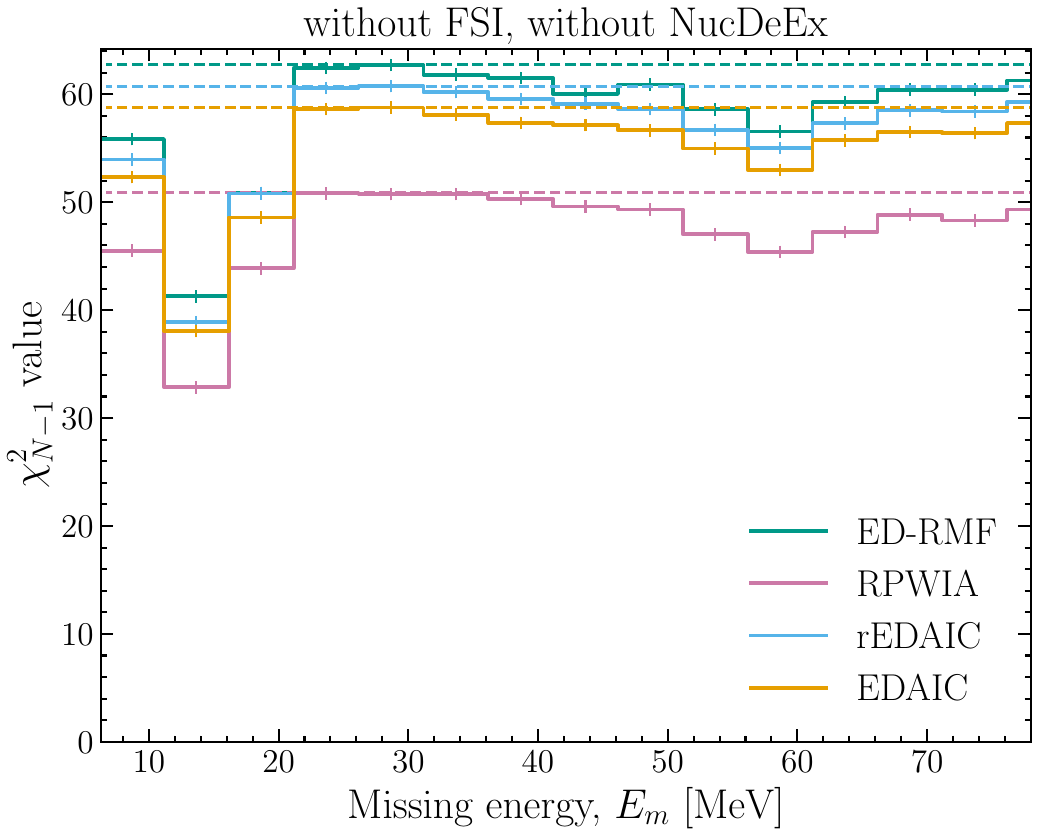}
    \end{minipage}\hfill
    \begin{minipage}{0.32\textwidth}
        \centering
        \includegraphics[width=\linewidth]{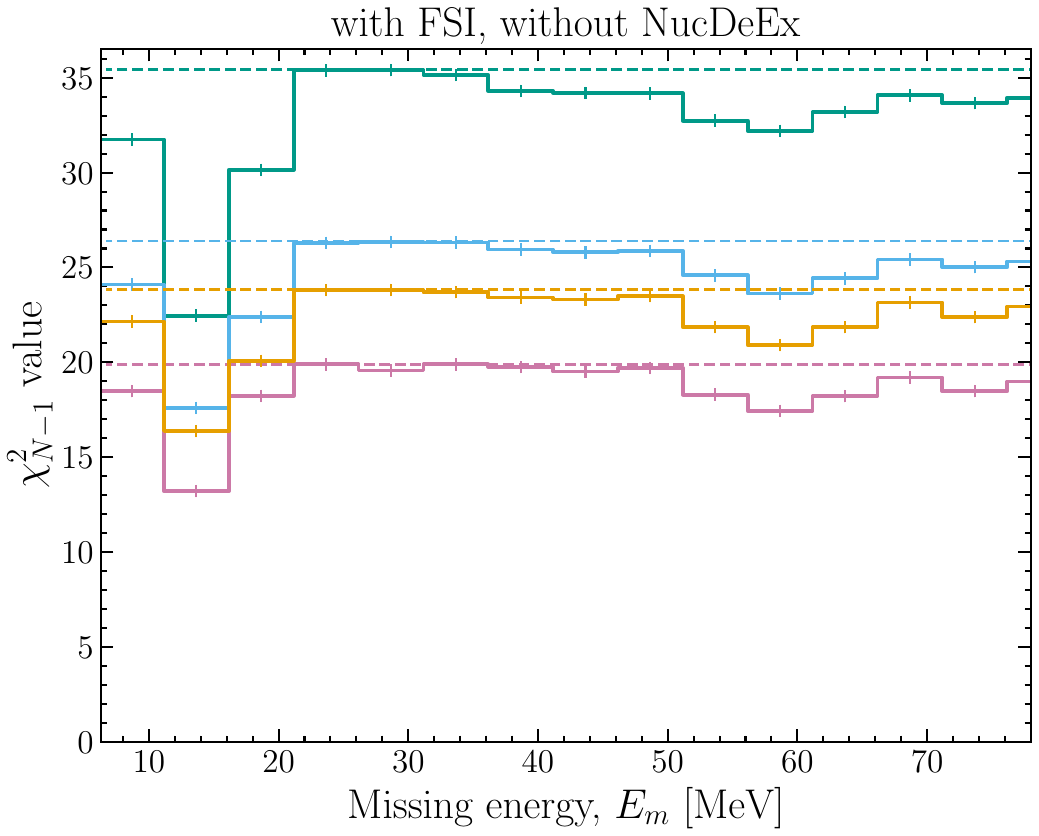}
    \end{minipage}\hfill
    \begin{minipage}{0.32\textwidth}
        \centering
        \includegraphics[width=\linewidth]{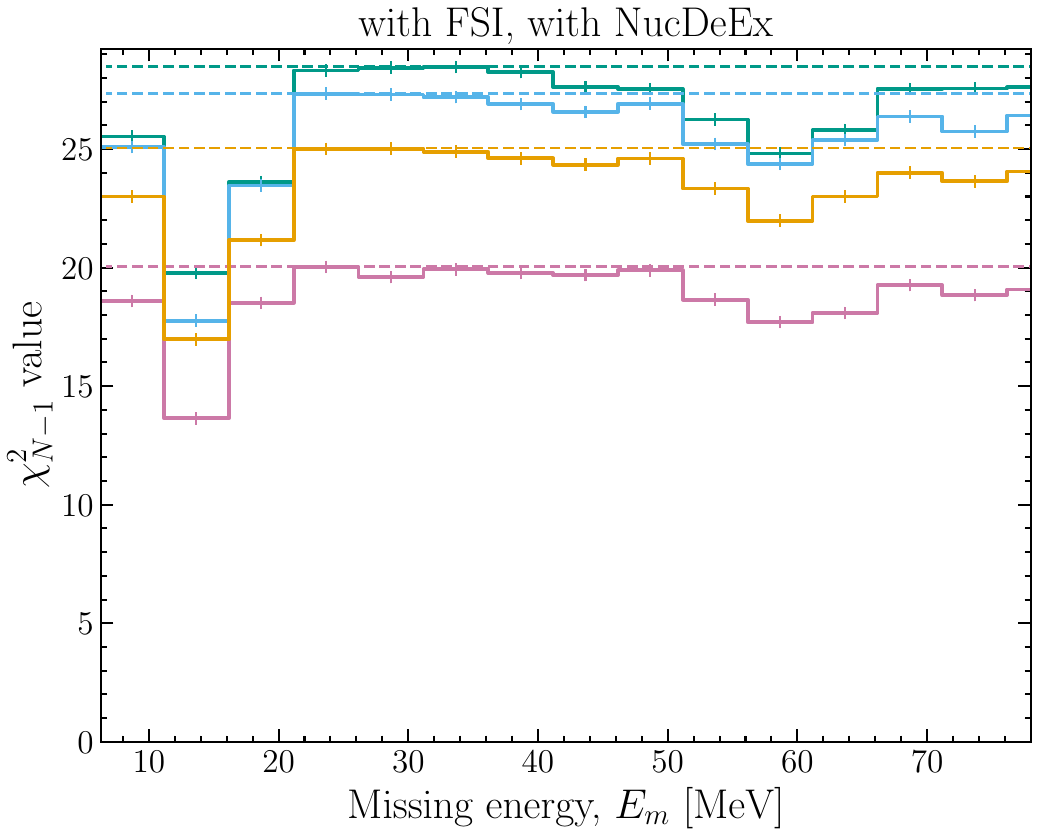}
    \end{minipage}
    \caption{The same as Figure~\ref{fig:chi2_Nminus1_wo_1NKO} but showing the ED-RMF, RPWIA, EDAI and rEDAI nuclear potentials.}
    \label{fig:chi2_missE_NucPots_wo_1NKO}
\end{figure*}
\begin{figure*}[htbp]
    \centering
    \begin{minipage}{0.32\textwidth}
        \centering
        \includegraphics[width=\linewidth]{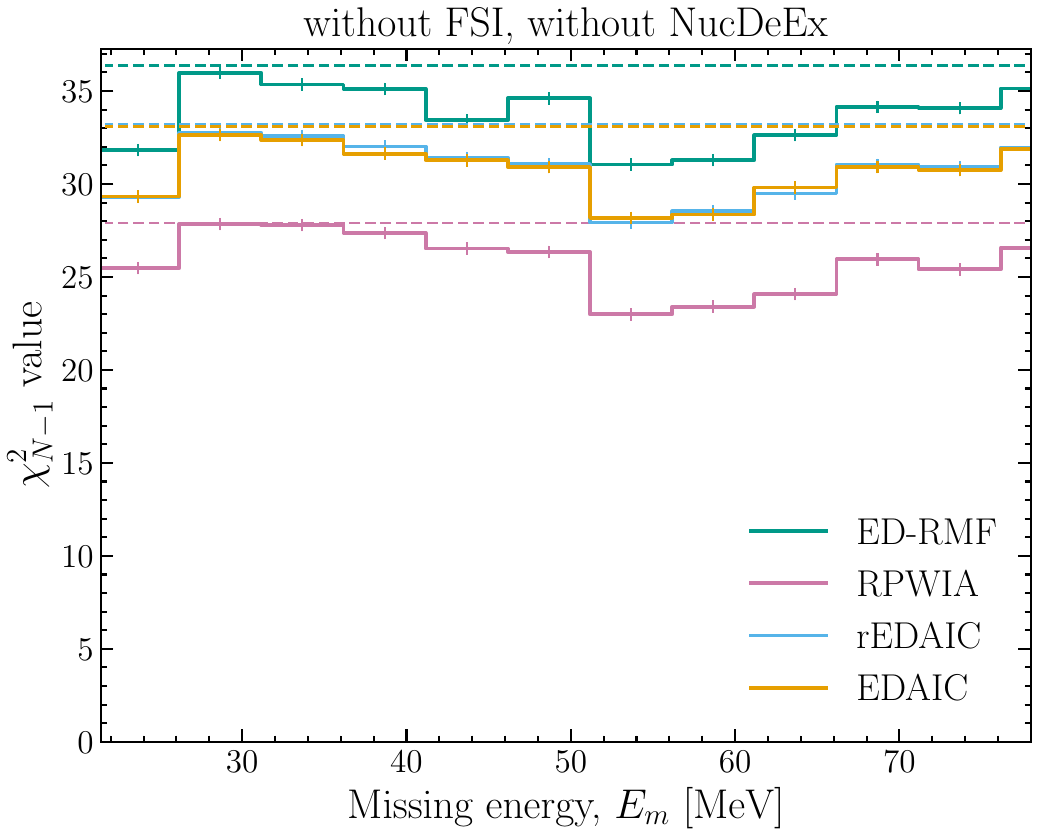}
    \end{minipage}\hfill
    \begin{minipage}{0.32\textwidth}
        \centering
        \includegraphics[width=\linewidth]{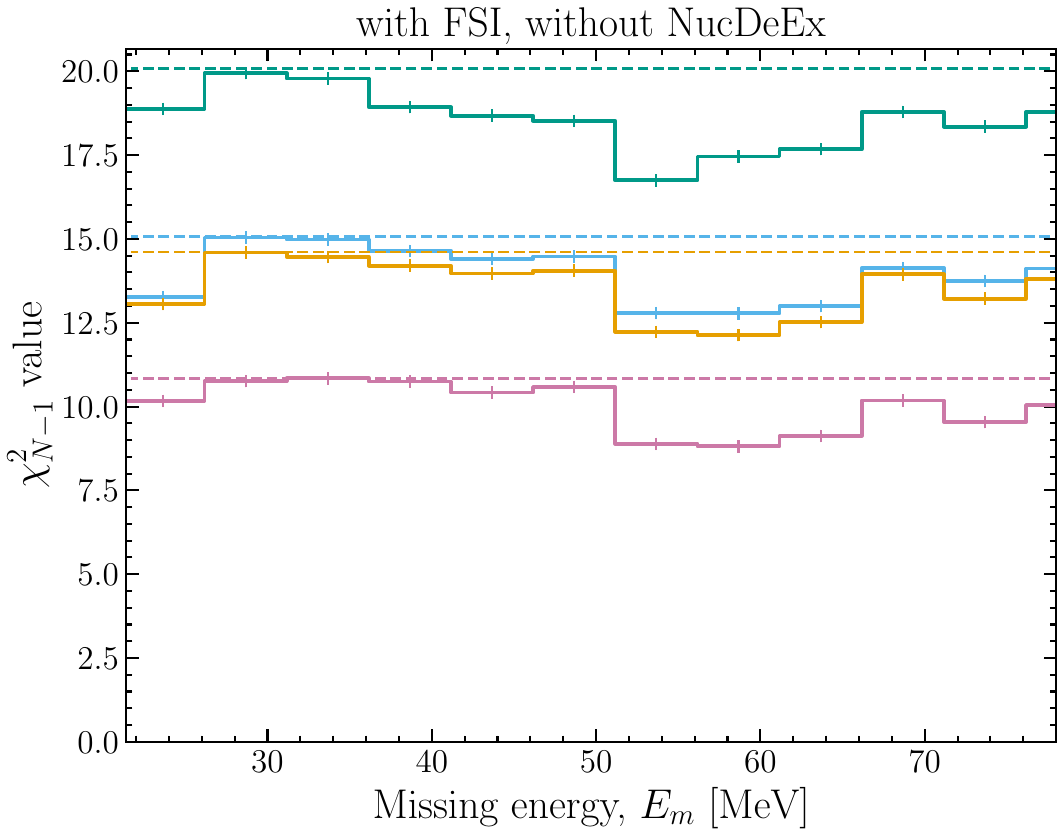}
    \end{minipage}\hfill
    \begin{minipage}{0.32\textwidth}
        \centering
        \includegraphics[width=\linewidth]{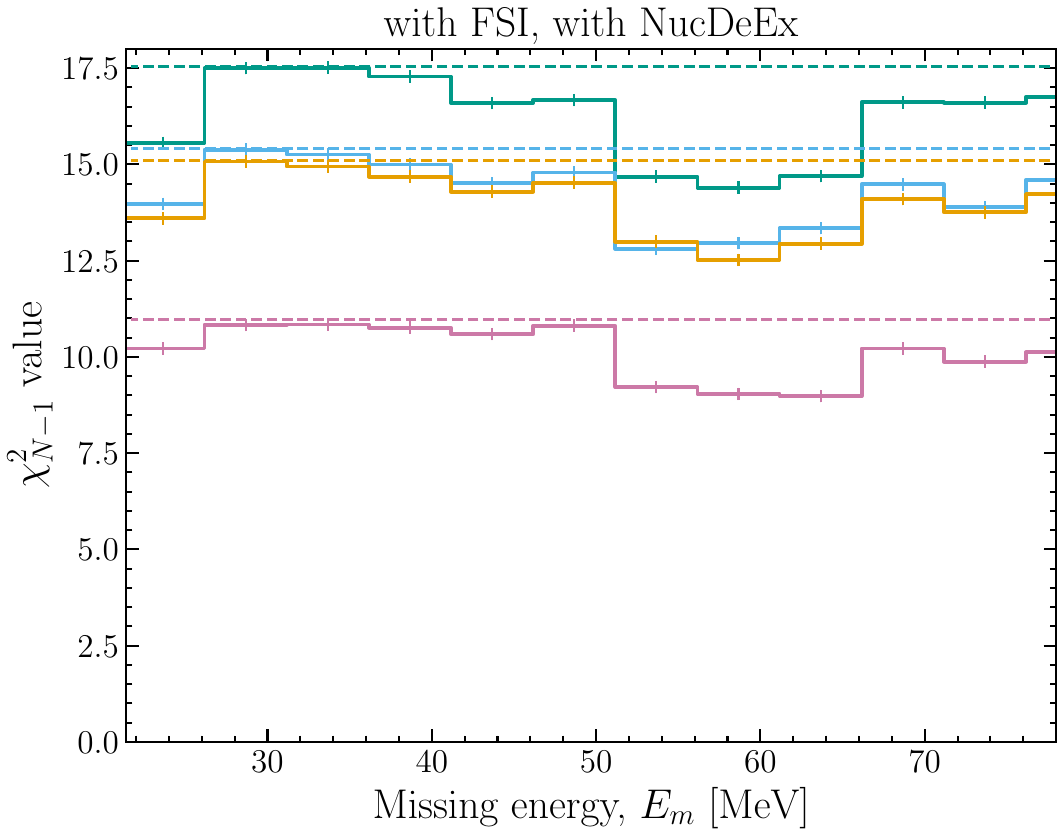}
    \end{minipage}
    \caption{The same as Figure~\ref{fig:chi2_missE_NucPots_wo_1NKO} but showing only the bins above the single nucleon knockout threshold.}
    \label{fig:chi2_missE_NucPots_w_1NKO}
\end{figure*}

From the left sides of Figures~\ref{fig:missE_NucPots},~\ref{fig:chi2_missE_NucPots_wo_1NKO} and~\ref{fig:chi2_missE_NucPots_w_1NKO}, the contributions to the $\chi^{2}$ across all bins behave similar in a qualitative way. The lowest $\chi^{2}$ values are obtained by the RPWIA model, largely because of its reduced contribution from the $1p_{3/2}$ peak and increased strength in the $1s_{1/2}$ peak. Still, without the effect of the inelastic FSI cascade and nuclear de-excitations, all models fail the $p$-value test, notably due to insufficient strength at high $E_m$ values.\par
Qualitatively, turning on the NEUT cascade and de-excitation routine has the same effect as described in Section~\ref{sec:results_discussion}. Quantitatively, however, the the RMF-based models are now past the $p$-value threshold with the inclusion of the NEUT cascade when other potentials are used, even before applying the 1NKO threshold. This is the case for the RPWIA, EDAI and rEDAI potentials, as showcased in Table~\ref{tab:DiffPots_Em_profile_config_chi2}. The inclusion of the 1NKO threshold leads all models to pass the $p$-value test. \par
When switching on the nuclear de-excitation routine, the $\chi^{2}$ values improve marginally for the ED-RMF model, and worsen slightly for the other models, both with and without the 1NKO threshold. This change causes the rEDAI potential to migrate below the $p$-value limit without the inclusion of the 1NKO threshold.\par

Throughout the comparisons, the RPWIA potential is consistently lower in the $1p_{3/2}$ peak in comparison to the other potentials. This results in a significantly improved $\chi^{2}$ value. With the inclusion of the 1NKO threshold, no model is rejected based on the obtained $p$-values.

\begin{table} [htbp]
    \caption{$\chi^{2}$ and $p$-values for the ED-RMF, RPWIA, EDAI and rEDAI nuclear potentials in NEUT. The values computed with configurations including the NEUT cascade (given by the FSI column), nuclear deexcitation and single nucleon knockout threshold (given by the 1NKO column) are shown for each configuration. Rows highlighted in bold font are not rejected given the $p$-values. The $p$-values are rounded to two significant figures.}
    \label{tab:DiffPots_Em_profile_config_chi2}
    \centering
    \begin{tabular*}{1.0\columnwidth}{@{\extracolsep{\fill}}c c c c c c }
    \hline \hline
    Nuclear potential & FSI & Deexcitation & 1NKO & $\chi^{2}/N_{\text{d.o.f}}$ & $p$-value\\ [0.5 ex]
    \hline
    ED-RMF & No & No & No & 62.72/16 & 0.00 \\
    RPWIA & No & No & No & 50.86/16 &  0.00\\
    EDAI & No & No & No & 58.79/16 & 0.00\\
    rEDAI & No & No & No & 60.75/16 & 0.00\\
    \hline
    ED-RMF & Yes & No & No & 35.41/16 & 0.00 \\
    \textbf{RPWIA} & \textbf{Yes} & \textbf{No} & \textbf{No} & $\mathbf{19.89/16}$ &  $\mathbf{0.22}$\\
    \textbf{EDAI} & \textbf{Yes} & \textbf{No} & \textbf{No} & $\mathbf{23.85/16}$ & $\mathbf{0.09}$\\
    \textbf{rEDAI} & \textbf{Yes} & \textbf{No} & \textbf{No} & $\mathbf{26.39/16}$ & $\mathbf{0.05}$\\
    \hline
    ED-RMF & Yes & Yes & No & 28.48/16 & 0.03 \\
    \textbf{RPWIA} & \textbf{Yes} & \textbf{Yes} & \textbf{No} & $\mathbf{20.04/16}$ &  $\mathbf{0.22}$\\
    \textbf{EDAI} & \textbf{Yes} & \textbf{Yes} & \textbf{No} & $\mathbf{25.05/16}$ & $\mathbf{0.07}$\\
    rEDAI & Yes & Yes & No & 27.34/16 & 0.04\\
    \hline
    ED-RMF & No & No & Yes & 36.37/13 & 0.00 \\
    RPWIA & No & No & Yes & 27.90/13 &  0.01\\
    EDAI & No & No & Yes & 33.11/13 & 0.00\\
    rEDAI & No & No & Yes & 33.20/13 & 0.00\\
    \hline
    \textbf{ED-RMF} & \textbf{Yes} & \textbf{No} & \textbf{Yes} & $\mathbf{20.07/13}$ & $\mathbf{0.09}$ \\
    \textbf{RPWIA} & \textbf{Yes} & \textbf{No} & \textbf{Yes} & $\mathbf{10.84/13}$ &  $\mathbf{0.62}$\\
    \textbf{EDAI} & \textbf{Yes} & \textbf{No} & \textbf{Yes} & $\mathbf{14.61/13}$ & $\mathbf{0.33}$\\
    \textbf{rEDAI} & \textbf{Yes} & \textbf{No} & \textbf{Yes} & $\mathbf{15.06/13}$ & $\mathbf{0.31}$\\
    \hline
    \textbf{ED-RMF} & \textbf{Yes} & \textbf{Yes} & \textbf{Yes} & $\mathbf{17.54/13}$ & $\mathbf{0.18}$ \\
    \textbf{RPWIA} & \textbf{Yes} & \textbf{Yes} & \textbf{Yes} & $\mathbf{10.97/13}$ &  $\mathbf{0.61}$\\
    \textbf{EDAI} & \textbf{Yes} & \textbf{Yes} & \textbf{Yes} & $\mathbf{15.10/13}$ & $\mathbf{0.30}$\\
    \textbf{rEDAI} & \textbf{Yes} & \textbf{Yes} & \textbf{Yes} & $\mathbf{15.40/13}$ & $\mathbf{0.28}$\\
    \hline \hline
    \end{tabular*}
\end{table}

\section{Prospects for improving the RMF nuclear model}\label{sec:ap:RMF-Em-profile}
\begin{figure}[htbp]
    \centering
    \includegraphics[width=1.0\columnwidth]{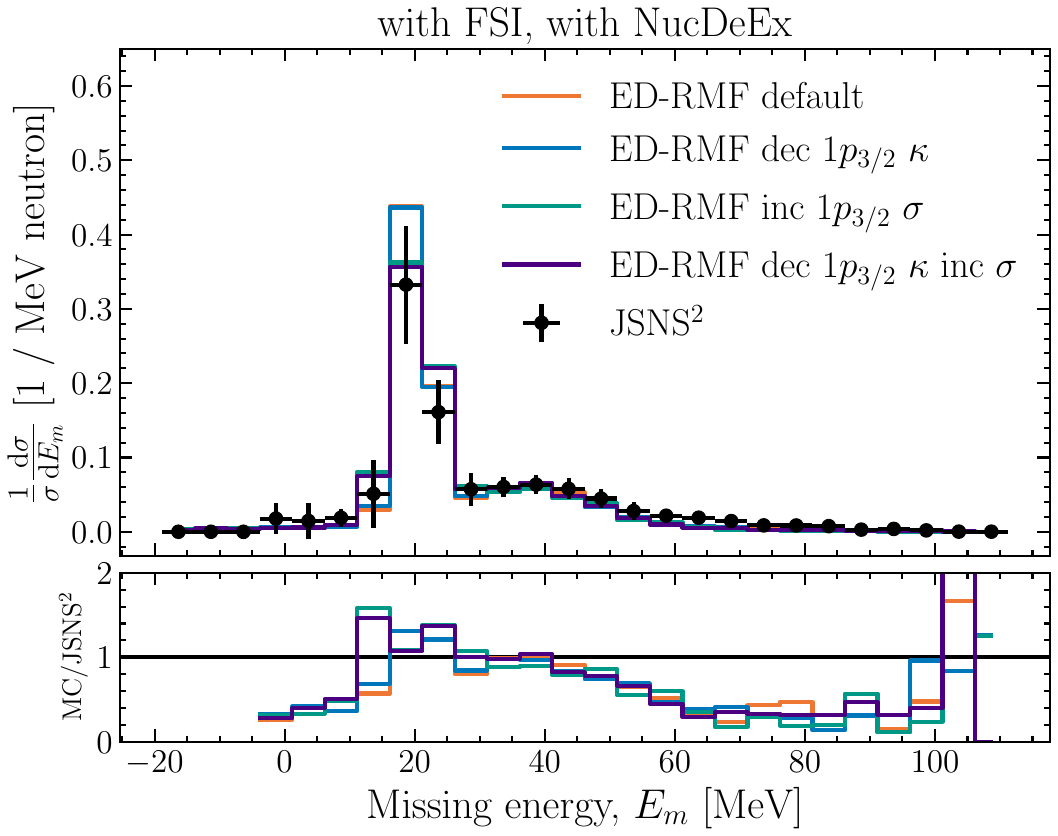}
    \caption{The normalised missing energy differential cross section ED-RMF model with changes to the modeling of the ground state. Both the NEUT cascade and \textsc{NucDeEx} are switched on.
    The single nucleon knockout threshold is not applied. The data points are taken from JSNS$^2$ measurement~\cite{KDAR:Marzec:2025}.}
    \label{fig:FSI_DeEx_change_RMF}
\end{figure}

Because the ED-RMF model uses Gaussians to model the nuclear shells, they can be easily altered. In this appendix, the ground-state shell was altered by increasing the width of the shell from 2\,MeV to 3\,MeV, decreasing the occupancy of the shell from $3.3/4.0$ to $3.0/4.0$ and performing both at the same time. In the case where the occupancy was decreased, the $1s_{1/2}$ shell occupancy was increased in order to keep the total number of nucleons constant. Figure~\ref{fig:FSI_DeEx_change_RMF} shows the comparison with the measurement when changing the missing energy profile. The effect of the modeling changes on the $\chi^{2}$ can be seen in Table~\ref{tab:Em_profile_config_chi2} both with and without the 1NKO threshold. In all cases without the 1NKO, the ED-RMF model is still rejected and is not improved significantly. With the 1NKO threshold applied, the $\chi^{2}$ is improved in the case where only the ground-state occupancy is decreased; all other changes appear to worsen the agreement with the measurement. 

\begin{table*} [htbp]
    \caption{$\chi^{2}$ and $p$-values for the ED-RMF model for the cases where the Gaussian description of the ground-state shell is altered by changing the shell occupancy and width. Both with and without the single nucleon knockout threshold are shown. The rows highlighted in bold are not rejected given the $p$-values. The $p$-values are rounded to two significant figures.}
    \label{tab:Em_profile_config_chi2}
    \centering
    \begin{tabular*}{0.9\textwidth}{@{\extracolsep{\fill}}c c c c c c }
    \hline \hline
    Change & 1NKO threshold & FSI & Deexcitation & $\chi^{2}/N_{\text{d.o.f}}$ & $p$-value\\ [0.5 ex]
    \hline
    $1p_{3/2}$ $\sigma$ increase & No& Yes & Yes & 32.95/16 & 0.00 \\
    $1p_{3/2}$ shell occupancy decrease & No & Yes & Yes & 30.49/16 & 0.02\\
    $1p_{3/2}$ $\sigma$ increase and decrease occupancy & No & Yes & Yes & 28.65/16 & 0.03\\
    \hline
    $1p_{3/2}$ $\sigma$ increase & Yes& Yes & Yes & 23.32/13 & 0.04 \\
    $\mathbf{1p_{3/2}}$ \textbf{shell occupancy decrease} & \textbf{Yes} & \textbf{Yes} & \textbf{Yes} & \textbf{17.21/13} & \textbf{0.19}\\
    $\mathbf{1p_{3/2}}$ $\mathbf{\sigma}$ \textbf{increase and decrease occupancy} & \textbf{Yes} & \textbf{Yes} & \textbf{Yes} & \textbf{21.37/13} & \textbf{0.07}\\
    \hline \hline
    \end{tabular*}
\end{table*}

\section{Changing the strength of the NEUT cascade}\label{sec:ap:FSI_strength}
\begin{figure*}[htbp]
    \centering
    \begin{minipage}{0.32\textwidth}
        \centering
        \includegraphics[width=\linewidth]{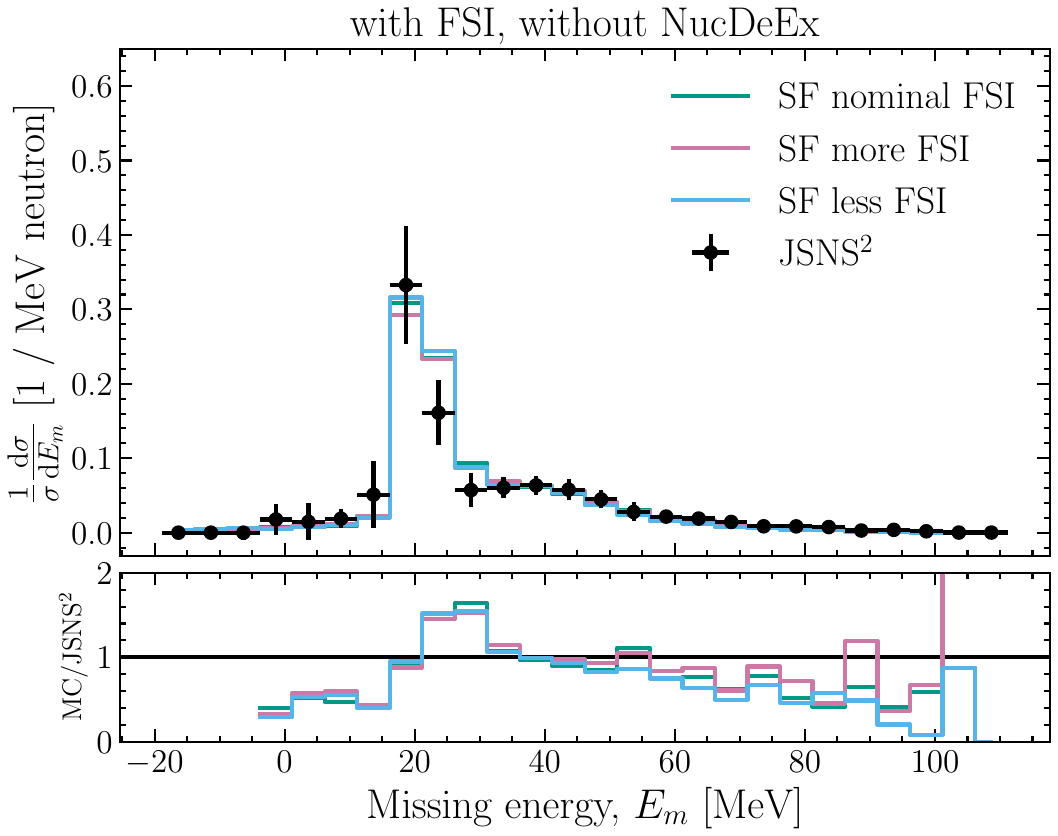}
    \end{minipage}\hfill
    \begin{minipage}{0.32\textwidth}
        \centering
        \includegraphics[width=\linewidth]{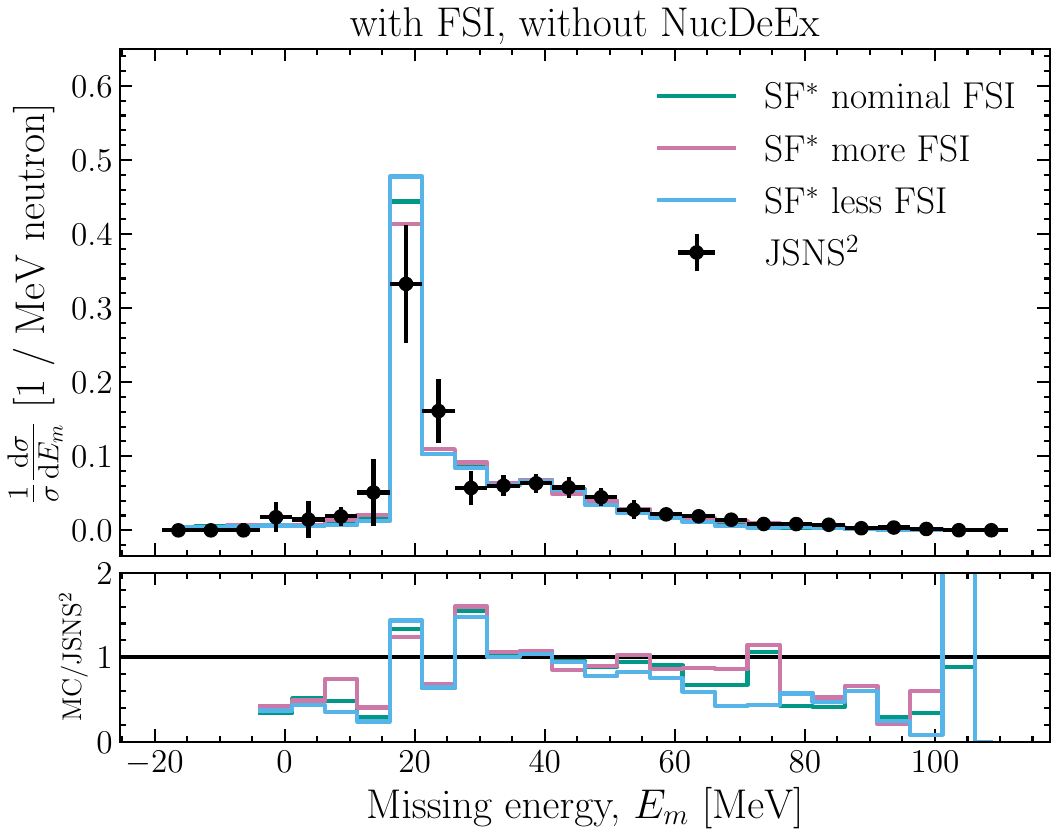}
    \end{minipage}\hfill
    \begin{minipage}{0.32\textwidth}
        \centering
        \includegraphics[width=\linewidth]{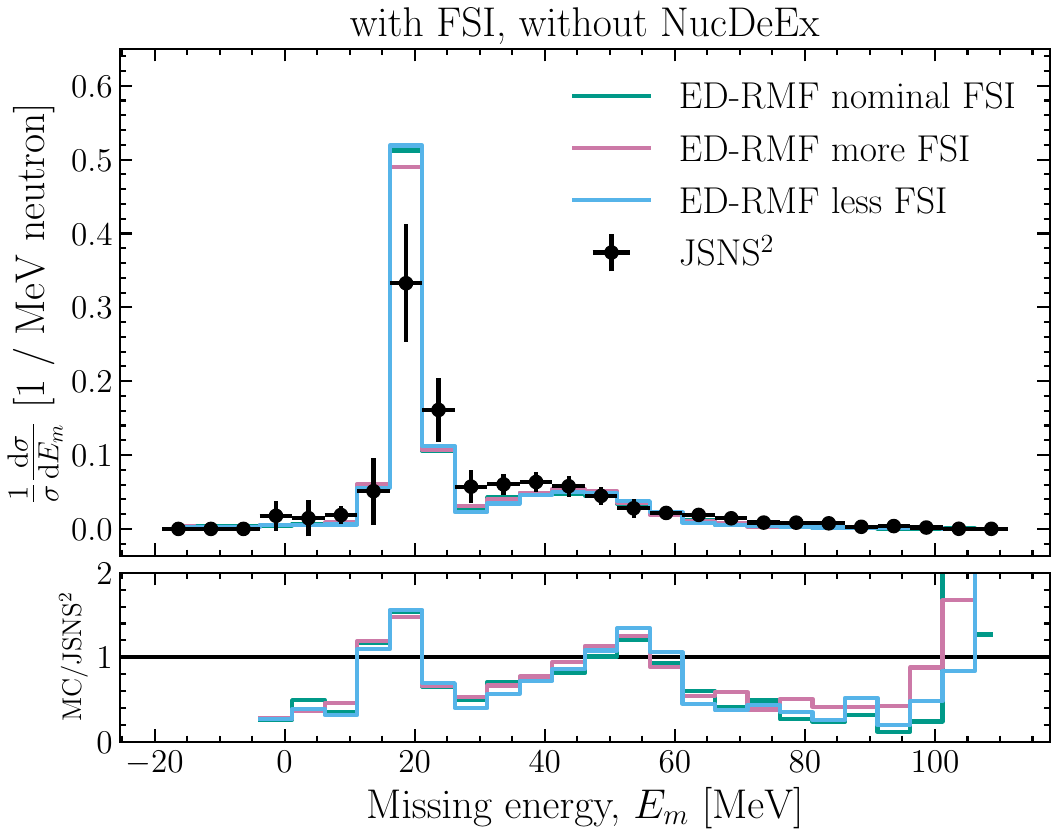}
    \end{minipage}
    \caption{The SF, SF$^{*}$ and ED-RMF models with changing strength of the NEUT FSI cascade. The FSI strength, determining the nucleon rescattering amount, is changed by $\pm 30$\%.}
    \label{fig:FSI_strength}
\end{figure*}

While this work uses only the NEUT generator's cascade model, it is informative to investigate how the strength of the applied nucleon rescattering FSI impacts the obtained results. Figure~\ref{fig:FSI_strength} shows the impact of changing the FSI strength from the cascade. The $\chi^{2}$ and $p$-values are given in Table~\ref{tab:FSI_strength}. The FSI strength is increased and decreased by 30\% from the nominal NEUT value\footnote{The 30\% value is derived from Ref.~\cite{Niewczas:2019fro}.}, which corresponds to a decrease and increase of the probability of FSI, respectively. In the case of the SF model, increasing the strength of the FSI yields a significant improvement of the $\chi^{2}$ and $p$-value, whereas decreasing it results in a slight worsening in the agreement with the measurement. For the SF$^{*}$ model, the $\chi^2$ values improve (worsen) when increasing (decreasing) the FSI strength, but all variations remain rejected at the $p$-value threshold. The effect is similar for the ED-RMF model, with slightly better $\chi^2$ values compared to those for the SF$^{*}$ model. In general, we find that the measurements favors increasing the strength of the FSI in all cases. In the case of the SF model, agreement is improved in both the $1p_{3/2}$ and the $1s_{1/2}$ shell peaks, as well as the transition between the $1p_{3/2}$ and the $1s_{1/2}$ peaks and the tail of the distribution. Since the ED-RMF model already overestimates the $1p_{3/2}$, the increase in FSI strength helps this agreement, but not as much as that of the SF$^{*}$ model, in which the peak now becomes consistent with the measurement. Other comparisons of the NEUT FSI cascade to proton scattering measurements, such as those presented in Ref.~\cite{Dytman:2021}, find a similar preference towards increasing the FSI strength in NEUT.
The study presented in this section only varies a single degree of freedom within the NEUT cascade, and does not represent an exhaustive examination of the effects of FSI modeling on the agreement with the measurement. A similar analysis to the one presented in this text, but using the NuWro~\cite{NuWro} event generator, reports that the NuWro cascade actually worsens the agreement of the ED-RMF model (with an improved missing energy profile)~\cite{Franco-Patino26}, despite having a similar description of transparency to that of NEUT. This effect is in conjunction with other different choices for nuclear effects and so the direct impact of the NuWro cascade is difficult to isolate. A dedicated study investigating the effects of other FSI cascades, such as the time-like Achilles~\cite{Achilles:PhysRevD.107.033007} cascade, is required in order to fully characterize the impact of FSI modeling.

\begin{table*} [htbp]
    \caption{$\chi^{2}$ and $p$-values for the SF, SF${*}$ and ED-RMF models when changing the strength of the NEUT FSI cascade by $\pm 30$\%. All samples are generated with \textsc{NucDeEx} off. The rows highlighted in bold are not rejected given the $p$-values. The $p$-values are rounded to two significant figures.}
    \label{tab:FSI_strength}
    \centering
    \begin{tabular*}{0.9\textwidth}{@{\extracolsep{\fill}}c c c c c c }
    \hline \hline
    Model & FSI strength & $\chi^{2}/N_{\text{d.o.f}}$ & $p$-value\\ [0.5 ex]
    \hline
    \textbf{SF} &  \textbf{nominal} & \textbf{14.11/16} & \textbf{0.59} \\
    \textbf{SF} &  $\mathbf{+30}$\textbf{\%} & \textbf{9.36/16} & \textbf{0.90}\\
    \textbf{SF} &  $\mathbf{-30}$\textbf{\%} & \textbf{16.40/16} & \textbf{0.43}\\
    \hline
    SF$^{*}$ &  nominal & 44.53/16 & 0.00 \\
    SF$^{*}$ &  $+30$\% & 35.06/16 & 0.00\\
    SF$^{*}$ &  $-30$\% & 53.41/16 & 0.00\\
    \hline
    ED-RMF &  nominal & 43.04/16 & 0.00 \\
    ED-RMF &  $+30$\% & 34.35/16 & 0.01\\
    ED-RMF &  $-30$\% & 48.77/16 & 0.00\\
    \hline \hline
    \end{tabular*}
\end{table*}

\FloatBarrier
\bibliography{main}

\end{document}